\documentclass[conference]{IEEEtran}
\ifCLASSINFOpdf
\else
\fi
\usepackage{tikz}
\usepackage{amsmath}
\usepackage{filecontents}
\usepackage{graphicx}
\DeclareGraphicsExtensions{.pdf,.png,.jpg}
\usepackage{algorithmic}
\usepackage{textcomp}
\usepackage{xcolor}
\usepackage{color}
\usepackage{booktabs}
\usepackage{array}
\usepackage{pifont}
\usepackage{cleveref}
\usepackage{subfigure}
\graphicspath{{./figure/}}
\usepackage{url}
\usepackage{textcomp}
\usepackage[T1]{fontenc}

\usepackage{breakurl}
\usepackage[mathscr]{eucal}
\usepackage{algorithm}
\usepackage{algorithmic}
\usepackage{xspace}
\usepackage{multirow}
\usepackage{makecell}
\usepackage{booktabs}
\usepackage{diagbox}
\usepackage{threeparttable}
\usepackage{graphicx}
\usepackage{float}
\usepackage{ulem}
\usepackage{footnote}

\newcommand{\alias}{\texttt{ReThink}\xspace
}
\newcommand{\dos}{\texttt{DoS}\xspace}
\newcommand{\damage}{\texttt{Damage}\xspace}
\newcommand{\damp}{\texttt{Damping}\xspace}

\begin{document}
\title{\alias: Reveal the Threat of Electromagnetic Interference on Power Inverters}


%\author{
%    \IEEEauthorblockN{Fengchen Yang\IEEEauthorrefmark{$^*$}}
%	\IEEEauthorblockA{Zhejiang University; ZJU QI-ANXIN \\IoT Security Joint Labratory\\
%		yangfengchen@zju.edu.cn}\\	
%  \and
%  
%	\IEEEauthorblockN{Zihao Dan\IEEEauthorrefmark{$^*$}}
%	\IEEEauthorblockA{Zhejiang University; ZJU QI-ANXIN \\IoT Security Joint Labratory\\
%		danzihao@zju.edu.cn}\\
%  \and
%  
%	\IEEEauthorblockN{Kaikai Pan\IEEEauthorrefmark{$\dagger$}}
%	\IEEEauthorblockA{Zhejiang University; ZJU QI-ANXIN \\IoT Security Joint Labratory\\
%		pankaikai@zju.edu.cn}  \\
%    \and
%    
%	\IEEEauthorblockN{Chen Yan}
%	\IEEEauthorblockA{Zhejiang University; ZJU QI-ANXIN \\IoT Security Joint Labratory\\
%		yanchen@zju.edu.cn} \\
%  \and
%  
%	\IEEEauthorblockN{Xiaoyu Ji}
%	\IEEEauthorblockA{Zhejiang University; ZJU QI-ANXIN \\IoT Security Joint Labratory\\
%		xji@zju.edu.cn} \\
%    \and
%    
%	\IEEEauthorblockN{Wenyuan Xu}
%	\IEEEauthorblockA{Zhejiang University; ZJU QI-ANXIN \\IoT Security Joint Labratory\\
%		wyxu@zju.edu.cn}
%
%  }

\author{
    \IEEEauthorblockN{Fengchen Yang$^{1,2*}$, Zihao Dan$^{1,2*}$, Kaikai Pan$^{1,2\dagger}$, Chen Yan$^{1,2}$, Xiaoyu Ji$^{1,2}$, and Wenyuan Xu$^{1,2}$}
    \IEEEauthorblockA{$^1$ Zhejiang University}
    \IEEEauthorblockA{$^2$ ZJU QI-ANXIN IoT Security Joint Labratory}
    \IEEEauthorblockA{\{yangfengchen, danzihao, pankaikai, yanchen, xji, wyxu\}@zju.edu.cn}
}


%Telephone: (800) 555--1212, Fax: (888) 555--1212}


\IEEEoverridecommandlockouts
\makeatletter\def\@IEEEpubidpullup{6.5\baselineskip}\makeatother
\IEEEpubid{\parbox{\columnwidth}{
		Network and Distributed System Security (NDSS) Symposium 2025\\
		23-28 February 2025, San Diego, CA, USA\\
		ISBN 979-8-9894372-8-3\\
		https://dx.doi.org/10.14722/ndss.2025.23691\\
		www.ndss-symposium.org
}
\hspace{\columnsep}\makebox[\columnwidth]{}}

\maketitle
% !TEX root = ../Theremin.tex
% \clearpage
\begin{abstract}	  
% 新版本
With the boom of renewable energy sources (RES), the number of power inverters proliferates. Power inverters are the key electronic devices that transform the direct current (DC) power from RES to the alternating current (AC) power on the grids, and their security can affect the stable operation of RES and even power grids. This paper analyzes the security of photovoltaic (PV) inverters from the aspects of internal sensors since they serve as the 
foundation for safe power conversion. We discover that both the embedded current sensors and voltage sensors are vulnerable to electromagnetic interference (EMI) of 1 GHz or higher, despite electromagnetic compatibility (EMC) countermeasures. Such vulnerabilities can lead to incorrect measurements and deceiving the control algorithms, and we design \alias that could produce three types of consequences on PV inverters by emitting carefully crafted EMI, i.e., Denial of Service (DoS), damaging inverters physically or damping the power output. We successfully validate these consequences on 5 off-the-shelf PV inverters, and even in a real-world microgrid, by transmitting EMI signals at a distance of $100 \sim 150 \mathrm{cm}$ and a total power within $20  \, \mathrm{W}$. Our work aims to raise awareness of the security of power electronic devices of RES, as they represent an emerging Cyber-Physical attack surface to the future RES-dominated grid. Finally, to cope with such threats, we provide hardware and software-based countermeasures. 

\renewcommand{\thefootnote}{}
\footnotetext{$^*$ Fengchen and Zihao contributed equally to this work.}
\footnotetext{$^\dagger$ Kaikai Pan is the corresponding author.}

\end{abstract}

\IEEEpeerreviewmaketitle

% !TEX root = ../Theremin.tex
% \clearpage
\section{Introduction}

\begin{figure}[t]
	%\centerline{\includegraphics[width=8.9cm]{figure/intro.pdf}}%\vspace{-2mm}
	%\centerline{\includegraphics[width=8.9cm,height=3.4cm]{figure/intro.pdf}}
	%
	\centerline{\includegraphics[width=\linewidth]{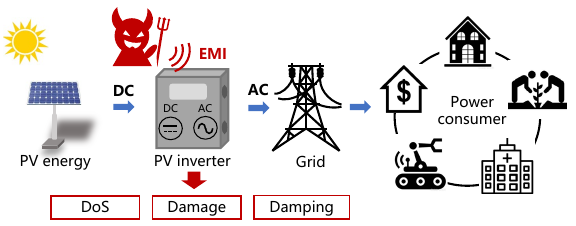}}
	
	\caption{An illustration of \alias: EMI can affect PV inverters and cause DoS or physical damage, or damping the power output.}
	\label{intro}%\vspace{-1mm}
\end{figure}

Renewable energy sources (RES), e.g., solar, wind, or hydroelectric power, are replacing fossil fuels to reduce 
their impact on global climate change~\cite{moosavian2013energy} and are expected to account for $47.7\%$ of all energy sources by 2040~\cite{panwar2011role}. As the penetration rate of RES continues to increase, it is critical to examine the emerging security issues of the power grids before RES constructions are finalized. Since most RES generates direct current (DC) power, yet the grids and power consumers operate on alternating current (AC) power, millions of power inverters have to be installed to convert DC power into AC power for each RES, as shown in Fig.~\ref{intro}. Thus, the security of power inverters can affect the smooth operation of RES power generation and even the stability of the power grids. Without loss of generality, we perform a systematic security analysis of solar inverter, aka., photovoltaic (PV) inverter, with the goal of providing security insights to device developers and designers, since PV is one of the most important RES and its capacity will produce about %1/3 
one-third of the world's annual energy consumption by 2050~\cite{9000G,3fenzhi1}.

In this paper, we focus on the distinct security of inverters, i.e., the security of analog sensors, since inverters rely on correct sensing of voltage and current of input power sources as well as the grids to ensure stable and safe power conversion. For instance, without accurate sensing of current and voltage, the inverter may fail to detect islanding conditions (when the grid is down but the inverter is still producing power) and potentially cause fire or electrify a maintenance technician~\cite{danger}. Sensors are known to be vulnerable to electromagnetic interference (EMI)~\cite{adc,pwm,ghost,trick}, and most PV inverters are installed in unguarded areas, e.g., resident backyards, building rooftops, or power plants in a desert~\cite{parida2011review}, whereby immersing sensors with malicious EMI signals is possible. 

These observations motivate us to perform further investigation into 
the impact of EMI on PV inverters, yet the DC-AC power conversion circuits inside inverters generally handle 50 watts up to 50 kilowatts~\cite{50w} and are a natural and strong source of EMI by design. For instance, power semiconductor switches that commutate at high switching frequencies will radiate EMI. Thus, all power inverters have to satisfy the electromagnetic compatibility (EMC) requirements by properly grounding, adding filters, and shielding so that they can operate normally in the presence of self and mutual interference. Although prior work~\cite{barua2020hall} has shown that static magnetic field can affect Hall sensors at a distance of 10 cm, it is unclear whether EMI injection could affect other types of embedded sensors, e.g., voltage sensors, and whether EMI signals can be crafted to precisely manipulate chosen sensors, as well as their consequences on inverters as a whole. 

After performing a systematic security analysis of the PV inverters on real inverters and microgrid~\footnote{Microgrid is a mini version of the grid, where it contains a group of interconnected loads and distributed energy resources and can connect to the grids or operate in an islanding mode~\cite{usdepartment, Robert}.}, we discover that both the embedded current and voltage sensors in PV inverters are vulnerable to EMI, \textit{although they conform to EMC standards on conduction and radiation interference}~\cite{emc2014}. We believe that such vulnerabilities are caused by three reasons. First, the EMC is designed to cope with unintentional interference, and its frequency band does not cover the range of effective EMI frequency. The EMC standard mainly considers two types of interference: the conducted interference in the range of $0.15  \, \mathrm{MHz}\sim 30  \, \mathrm{MHz}$ and the radiated interference in the range of $30  \, \mathrm{MHz}\sim 1  \, \mathrm{GHz}$~\cite{emc2014}. Yet, the effective EMI signals can be higher than $1  \, \mathrm{GHz}$. Second, although low-pass filters are meant to remove all interference signals with a frequency higher than $0.15 \,\mathrm{MHz}$, the real filters are not ideal and can let go of high-frequency signals~\cite{RFIFILTER}. 
Last but not least, %some unintentional 
inverter designs may unintentionally provide backdoors for EMI. For example, 
\ding{172} the LCD screen on the inverter creates a gap of EMC protection and a vulnerable window for EMI; 
\ding{173} the non-ideal printed circuit board~(PCB) alignment and device layout bring parasitic capacitance;
\ding{174} and the circuit's asymmetrical alignment on the PCB weakens its immunity to common-mode interference; 
\ding{175}~the control algorithms of PV inverters assume the reliability of sensor measurements and lack of consistency checking. Thus, false voltage and current measurements can trick the PV inverter's control algorithms.  
Currently, most medium-voltage power electronic converters still commonly suffer from parasitic capacitance~\cite{7907237}, and current research mainly focuses on predicting and reducing parasitic capacitance~\cite{qu2023parasitic, 7847438, 7519493, 8495188, 10107452}, but most methods will increase the material and manufacturing cost~\cite{7907237}.

%来呼吁安全研究者们重视光伏逆变器所面临的安全威胁
To illustrate the impact of the aforementioned vulnerabilities in combination, we design \textbf{\texttt{\alias}}\xspace(reveal the threat of EMI on inverters) that could produce three types of consequences on PV inverters by emitting carefully crafted EMI, as shown in Fig.~\ref{intro}. 

\begin{itemize}%[noitemsep]
	\vspace{-1mm}
\item \dos: The PV inverter shuts down completely, causing an instantaneous power reduction of PV generation to the grid or consumers. %to the grid and a grid AC frequency drops due to the imbalance between power consumption and generation.
\item \damage: The PV inverter can be physically burned out and has to be repaired or replaced. %It can also cause a frequency reduction. 
\item \damp: This type of threat causes the output power of PV inverters to be lower than their capability. Long-term continuous \damp will reduce the efficiency of the PV generation.
\end{itemize}

We have validated the consequences of \alias on a PV inverter development kit, 5 off-the-shelf kilowatt-level PV inverters, and a rural-scale microgrid operated in the real world, by transmitting EMI signals at a distance of $100 \sim 150  \, \mathrm{cm}$ and emission power within $20  \, \mathrm{W}$. Despite the fact that the power capabilities of PV inverters vary from a few kilowatts to %megawatts, 
60 kilowatts, the embedded current and voltage sensors operate on a voltage level of $5  \, \mathrm{V}$ and are all vulnerable to EMI signals. We have uploaded video demonstrations to the link \footnote{https://tinyurl.com/ReThinkDemoVideos\label{video}}. 
To enhance the security of PV inverters, we analyze the underlying causes of the vulnerabilities and propose hardware as well as software countermeasures, including blocking EMI transmissions, detecting measurement manipulation, and repairing control logic vulnerabilities. We hope these can provide guidelines for the design of the PV inverter, e.g., its sensor PCB and control algorithms.

To the best of our knowledge, this is the first systematic work analyzing the impact of EMI on PV inverters and validating on the real-world microgrid. Our work is complementary to existing studies on traditional software or communication-related issues, e.g., software vulnerabilities of inverters or DoS and replay attacks against DC microgrids~\cite{horus,benk,mo2009,wang2019,sahoo}. %That is, researchers have discovered 17 software vulnerabilities of PV inverters~\cite{horus}, problems associated with improper firmware upgrades inside solar plants~\cite{benk}, replay attacks~\cite{mo2009} %denial of service (DoS) attacks~\cite{wang2019}, and false data injection attacks (FDIA)~\cite{sahoo} against DC microgrids.  
The goal of our work is to raise awareness of the security of power electronic devices in the power grids as RES are increasingly being adopted and they represent an emerging CPS threat surface. We imagine that our analysis and conclusions may potentially lay the groundwork for analyzing other types of inverters and power electronic devices with similar sensors and control logic. In summary, our contributions are as follows:
% Hall current/voltage sensors and op-amp based voltage sensors.

\begin{itemize}%[noitemsep]
	\vspace{-1mm}
	\item We present a systematic security analysis of PV inverters and analyze the vulnerabilities of sensors and control algorithms susceptible to EMI signals.
	%\item We reveal a cost-effective EMI transmission link inside the PV inverter, and illustrate one-to-one and many-to-many control methods despite the difficulty of affecting one out of multiple sensors inside inverters.
	\item We illustrate the adversarial \alias scenarios that can shut down, permanently damage, and damp the power output of PV inverters, and we validate the threat on commercial PV inverters and a real-world microgrid.
	\item We analyze the root causes for the vulnerabilities and propose hardware and software countermeasures.
        
\end{itemize}

%Since GTIs share similar architectures, this paper illustrates the security implication of GTIs for PV power generation (we also note that PV power has become a competitive source because it has favorable cost properties). 

% !TEX root = ../Theremin.tex

\section{Background and Threat Model}
\label{sc2}
\subsection{Principle of PV Inverter}

PV inverters, like many other types of inverters, are the heart of every PV system. To satisfy various design requirements, PV inverters may have subtle differences in their circuit design~\cite{islam2015single}. After examining 47 inverters from three leading manufacturers~\cite{SungrowInverter, TmeicInverter, HuaweiInverter}, we found that 43 inverters employ a standard DC-DC-AC topology, and this predominant architecture is known as a Two-Stage Power Conversion (TSPC) system~\cite{dogga2019recent}, which is the focus of this paper. %Notably, 
Particularly, a PV inverter consists of a power conversion unit, multiple current and voltage sensors, and control algorithms. Since power generation efficiency is one of the most important goals, a PV inverter will track the PV panel's maximum power point (MPP) by sensing and incorporating various control algorithms to convert DC power into AC power. To understand the details, we introduce them below.
\begin{figure}[tb]
%	\vspace{-1em}
	\centerline{\includegraphics[width=8.5cm]{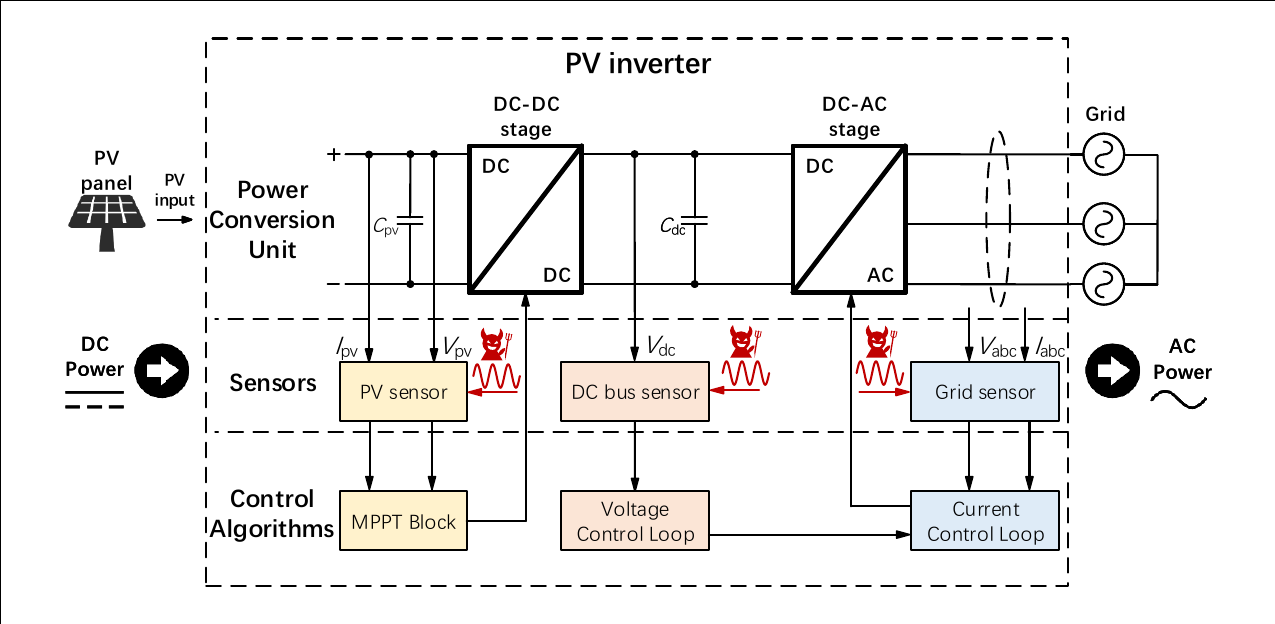}}
	\caption{A typical PV inverter can be modeled as a 3-layer structure: Power conversion unit-Sensor-Control algorithms.}
	\label{figPV}
%	\vspace{-1em}
\end{figure}
%The architecture of a typical PV inverter consists of a power conversion unit, sensors, and control algorithms. 

\subsubsection{Power Conversion Unit}
%\subsubsection{Structure of PV Inverter}
A typical TSPC PV inverter contains two parts: the DC-DC stage and the DC-AC stage, as shown in Fig.~\ref{figPV}.

\textbf{DC-DC Stage.} 
The primary function of the DC-DC stage is to increase the voltage level from the PV panel output, e.g., ranging from $30 \, \mathrm{V}$ to $60 \, \mathrm{V}$, to the one required by power grids, i.e., $325 \, \mathrm{V}$ peak for single-phase and $565 \, \mathrm{V}$ peak for three-phase.

\textbf{DC-AC Stage.} The DC-AC stage converts the direct current on the DC bus to the AC that can be fed into the grid through the inverter circuit, with the help of two control algorithms, i.e., voltage control loop and current control loop.

\subsubsection{Control Algorithm}
PV inverter relies on control algorithms to maintain the PV panels or arrays working at their maximum power state and convert DC into AC for integration into the grid. There are three main parts: the maximum power point tracking (MPPT) algorithm, the voltage control loop, and the current control loop. 

\textbf{MPPT Algorithm.} To maintain the highest energy conversion efficiency in various atmospheres~\cite{motahhir2020most,sarvi2022comprehensive}, the MPPT operates along a voltage-current (V-I) curve to identify the maximum power point~(MPP), where the V-I curve is an inherent characteristic of the PV panel and varies with the irradiance and temperature. The most commonly used MPPT algorithm is the Perturb and Observe (P\&O) method, where the basic idea is to try adding a perturbation to the inputs of PV inverters and measure the resulting power \cite{harrag2015variable}. 

\textbf{Voltage and Current Control Loop.} The role of the voltage control loop is to adjust the DC bus voltage $V_{dc}$ to a reference value. The DC bus capacitor functions as an energy buffer to stabilize the DC bus voltage. If the input power exceeds the output power, the capacitor $C_{dc}$ on the DC bus will continue to be charged, which will lead to an increase in $V_{dc}$ and trigger the voltage control loop to raise the output reference current $I_{dref}$, as shown in Fig.~\ref{VCL}. 
The function of the current control loop is to adjust the output current to match the output power of the PV inverter with its input power. Similar to the voltage control loop, it utilizes a PI controller to adjust the current based on the reference current provided by the voltage control loop.  Before entering the PI control, the coordinate system transformations (Clarke \& Park) \cite{yang2009grid} are applied to the measured three-phase voltage and current.

\textbf{Protection Mechanism of PV Inverter.}
% https://baijiahao.baidu.com/s?id=1655662687054247582&wfr=spider&for=pc
%In 
In the operation of PV inverters, a %range 
set of self-protection mechanisms are incorporated to prevent safety issues %caused by 
that may arise from device damage and circuit failure. The mechanisms considered in this paper include DC bus over-voltage protection, as well as AC over and under-voltage protection \cite{QSG}.

$\bullet$ DC bus over-voltage protection. The PV inverter continuously monitors the voltage of the DC bus. If the DC voltage exceeds a predefined threshold several times, the inverter disconnects from the grid and stops power generation.

$\bullet$ AC over and under voltage protection. When the inverter's output voltage is detected to be higher than the threshold range, it will disconnect itself from the grid. If the output voltage drops outside the allowable range of low voltage crossing~(20\%), the low voltage crossing function will activate, triggering an alarm. If the inverter's output voltage does not recover within a specified time, it will disconnect itself from the grid and stop working.

\subsection{Sensors of PV Inverter}
As illustrated in Fig.~\ref{figPV}, PV inverters rely on embedded sensors to measure voltage and current and feed them back to the control loop.

\begin{figure}[t]
	%\vspace{-1em}
	\centering
	\subfigure[Non-Hall voltage sensor.]{\includegraphics[width=4cm]{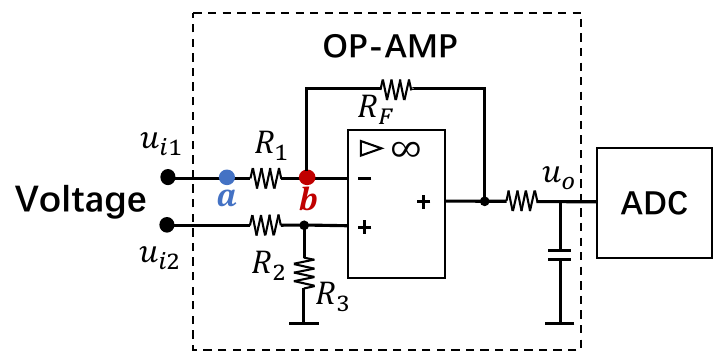}\label{vol}}
        \hspace{4mm}
	\subfigure[Hall current sensor.]{\includegraphics[width=4.2cm]{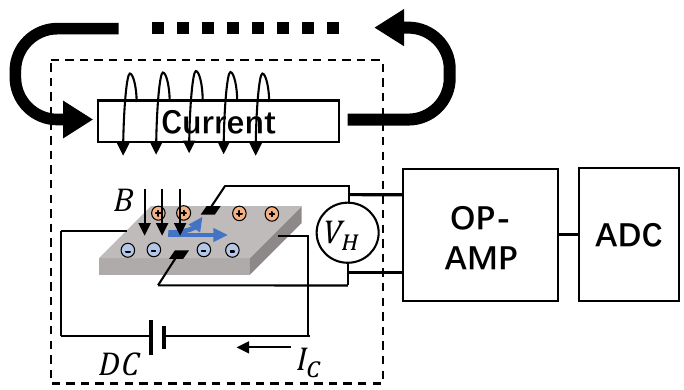}\label{hall}}
	\caption{The schematic of voltage and current sensors in the PV inverter~\cite{PCB}. The voltage sensor mainly comprises a differential op-amp circuit; the current sensor comprises a Hall chip and a differential op-amp circuit.}
	\vspace{-1em}
\end{figure}

\subsubsection{Non-Hall Voltage Sensor}\label{subsecvolsensor}
Voltage is one of the most important variables in a PV inverter. To detect the voltage in various control loops, voltage sensors convert hundreds of volts into a few volts that the analog-to-digital conversion~(ADC) module can handle. Besides, since inverters operate in complex electromagnetic environments and tend to generate common mode noise in the circuits, differential operational amplifiers~(op-amp) are often employed to suppress noises \cite{sbstract}. A typical structure of a differential op-amp circuit is shown in Fig.~\ref{vol}, and the magnification can be expressed as \Cref{A}: 
\begin{equation} 
	u_o = \frac{R_3\cdot(R_1 + R_F)}{R_1\cdot(R_2 + R_3)}\cdot u_{i2} - \frac{R_F}{R_1}\cdot u_{i1}
	\label{A} 
\end{equation} 

The magnification is determined by the resistors of the op-amp. In practice, resistors $R_1$ and $R_2$ usually consist of multiple divider resistors in series, and they step down the high voltage to a low voltage signal within $5  \, \mathrm{V}$; \textit{thus, for inverters from a few kilowatts to hundreds of kilowatts, the embedded voltage sensors shall be vulnerable to EMI signals at similar power levels.} The power levels are illustrated in Fig.~\ref{dp} of Evaluation.

\subsubsection{Hall Current Sensor}
\label{cs}
Since the current cannot be directly digitized by ADC modules, inverters typically use a Hall current sensor, which converts the magnetic field generated by the current into DC or AC voltage based on the Hall effect~\cite{7749133}. 

As shown in Fig.~\ref{hall}, the current $I$ generates a magnetic field $B$, and $B$ is proportional to $I$ according to Ampere's Law. 
Then the electrons moving on the electrode plate will be subjected to the Lorentz force $F_L$ in B and move to the sides of the electrode plate, and generate an electric field $E$ on the electrode plate. Finally, a balance state will be reached when the electric field force and the Lorentz force are equal, which can be formulated as \Cref{BE}, where $d$ is the width of the electrode plate and $q$ is the electrical charge. Since $B$ is proportional to $I$ and $V_H$ is proportional to $B$, the Hall sensor's output $V_H$ is proportional to the current $I$. Finally, Hall current sensors use a similar op-amp to suppress the common-mode noise in $V_H$ and output the measurement result. 
\begin{equation}
    F_L = F_E \Rightarrow B\cdot q\cdot v = q\cdot E = q\cdot \frac{V_H}{d}
    \label{BE}
\end{equation}

\subsection{Threat model}
\label{subsecthreatmodel}
We make the following assumptions about the adversary:

\textbf{Attack Goal.} The attacker's goal is to covertly cause the shutdown, power reduction, or even burnout of a PV inverter. Though ambitious attackers may target a group of inverters and try to create potentially escalated impacts such as voltage or frequency fluctuations or even blackouts in a local microgrid, we focus on basic attacks against individual inverters in this paper.

\textbf{Non-contact Access.}
We assume the attacker can approach the target inverters within a few meters, but she cannot physically touch or damage them due to safety and stealthiness concerns. Alternatively, the adversary can leave a camouflaged EMI device nearby and control it remotely.

\textbf{Prior Knowledge.} We assume that adversaries could have prior knowledge of the target inverter. Given that many PV inverters are commercial products readily available on the market, the adversary could acquire a PV inverter of the same model and conduct necessary tests beforehand. More favorably, in practice, PV systems in a region often use the same model of PV inverters.

% !TEX root = ../Theremin.tex
\section{Understanding the Impact of EMI on Embedded Sensors of PV Inverters}
\label{sc4}

%------------------------------------------------------------------------------
Sensors provide an entrance for EMI to impact PV inverters. In this section, we explore how EMI affects embedded voltage and current sensors of PV inverters through theoretical analysis and feasibility experiments.
%------------------------------------------------------------------------------
\subsection{Analysis of the EMI Impact on Sensors}\label{sc4.1.1}

\begin{figure}[t]
	\centering
	\subfigure[Transmission process of EMI signals in the voltage sensor.]{\includegraphics[width=5cm]{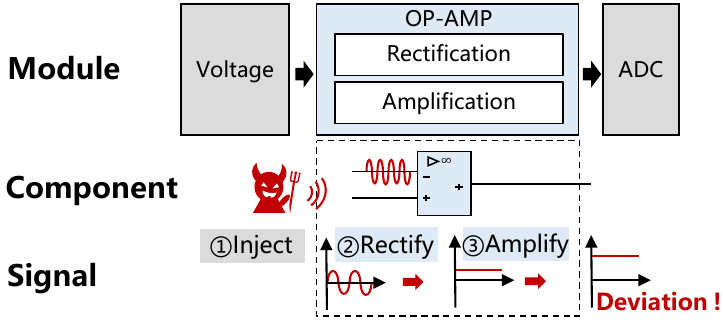}}\label{trans}
        \hspace{2mm}
	\subfigure[The parasitic capacitance of sensor's PCB.]{\includegraphics[width=3.3cm]{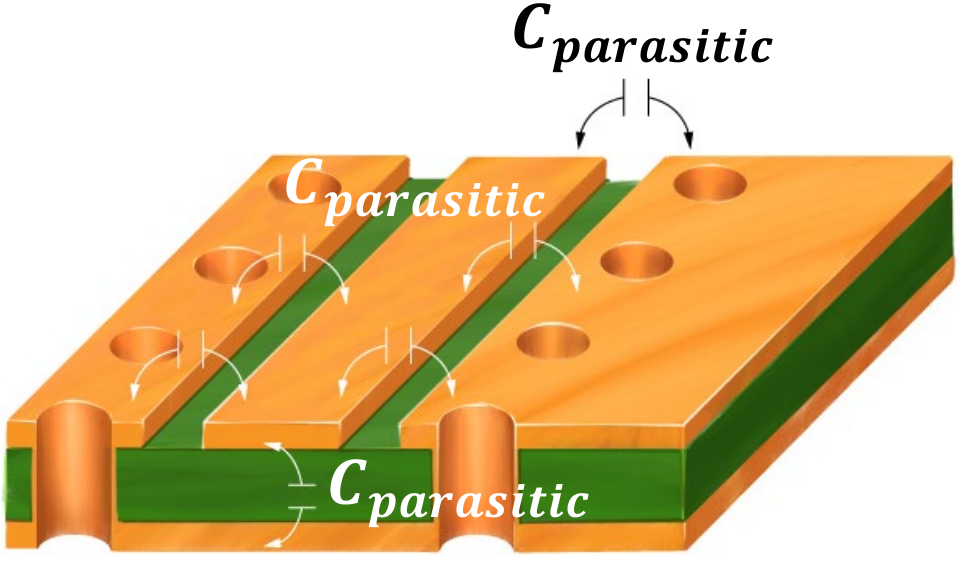}\label{stray}}
	\caption{The principle of EMI impact on voltage sensors. The EMI signal is coupled into the sensor circuit, and then rectified, amplified by the op-amp, and ultimately turned into an offset on the output.}
    \label{EMI_VOL}
\end{figure}

The key to exploring the impact of EMI on sensors is to identify the entry points and transmission path of EMI, while considering EMC measures. %into account. 
Given that the voltage sensor contains an op-amp circuit, while the current sensor contains not only an op-amp but also a Hall element, we analyze them separately.

\subsubsection{Impact of EMI on Voltage Sensors}
The sensor's %printed circuit board~(PCB) 
PCB usually carries parasitic capacitance and is susceptible to electromagnetic interference in the environment.
Besides, the op-amp circuit will further %``enhance'' 
rectify and amplify the coupled signals. The transmission process can be illustrated in Fig.~\ref{EMI_VOL}(a), and there are four steps:

\textbf{\textbullet~EMI signal injection.} Process \ding{172} in Fig.~\ref{EMI_VOL}(a) is EMI injection. Electromagnetic fields around the sensor can be injected into sensor circuits (e.g., input nodes) via electromagnetic coupling. Generally, according to the EMI transmission paths, EMI coupling methods can be divided into conductive coupling, inductive coupling, capacitive coupling, and radiative coupling~(also called radio frequency interference, RFI)~\cite{Cadence, Akanksha}. Among them, radiative coupling refers to the far-field coupling of higher-frequency signals in the microwave frequency range, which can be transmitted over longer distances. Notably, the conductors (e.g., copper wires and component pins) and the insulator (e.g., PCB substrate) on the sensor's PCB will form parasitic capacitance, as shown in Fig.~\ref{stray}. These parasitic capacitances are susceptible to the aforementioned high-frequency electric fields, which can introduce interfering signals. Thus, high-frequency EMI signals can be effectively coupled into the sensor circuits via radiative coupling. 
%To understand the potential electromagnetic coupling path of sensor circuits, we measured the reactance of the voltage sensor circuit to be 16.24~$\Omega$ at 10~kHz, making it an inductive circuit, which may be due to the parasitic inductance formed by the wires on the PCB. Thus a high-frequency magnetic field will interfere the sensor circuit via radiative coupling.}

\textbf{\textbullet~Nonlinear rectification effect.}
The amplifier can rectify the high-frequency AC signal at the input and generate a DC bias at the output. The main reason is that the bipolar junction transistor~(BJT) in the op-amp chip contains p-n junction diodes, which are efficient rectifiers due to their nonlinear current-voltage characteristics, especially in low-power op-amps~\cite{RFI}. When a high-frequency signal $v(t) = V_Xcos(2\pi f_Xt)$ is injected into the base-emitter junction of an op-amp BJT-based input stage, the output will generate an AC term $\Delta i_C(AC)$ at twice the input frequency and a DC term $\Delta i_C(DC)$~\cite{RFI}, which can be described by \Cref{icdc}:
\begin{equation}
		\Delta i_C(DC) = (\frac{V_X}{V_T})^2 \cdot \frac{I_C}{4}
		\label{icdc}
\end{equation}

\begin{figure}[t]
	\centerline{\includegraphics[width=8.5cm]{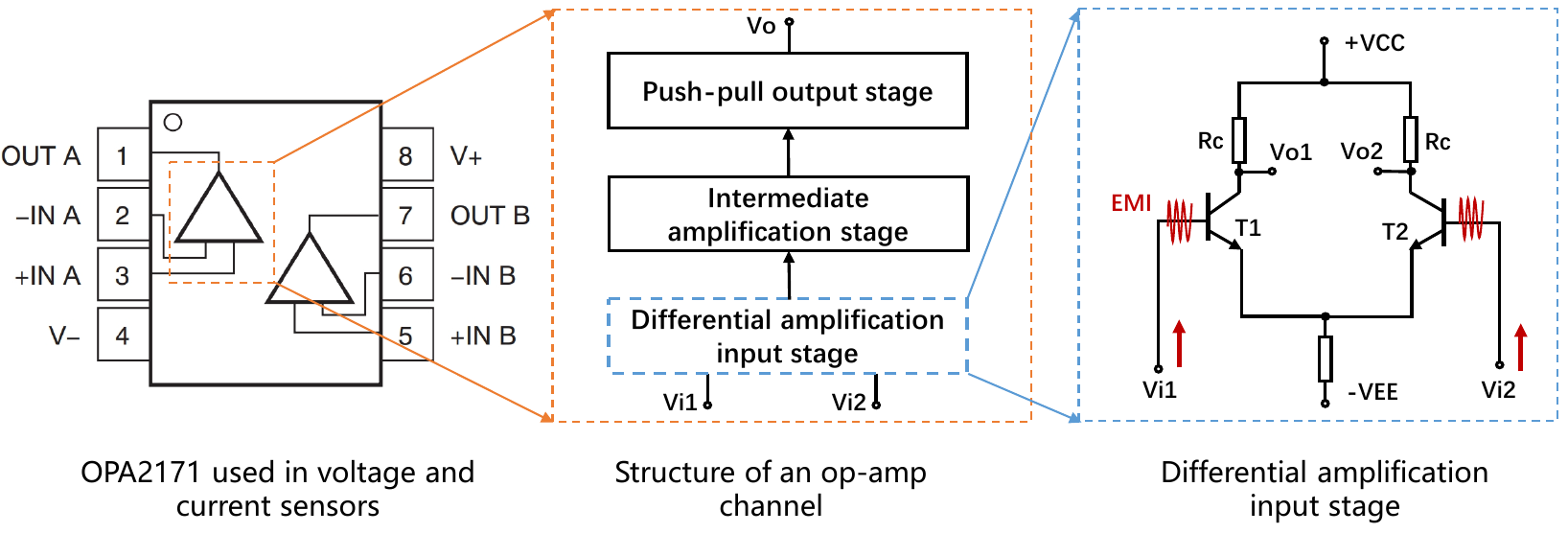}}%\vspace{-3mm}
	\caption{The structure of the
OPA2171 used in voltage and current sensors.}
	\label{chip}
\end{figure}

\textbf{\textbullet~Asymmetric differential effect.}
The asymmetric design of the op-amp circuit on the PCB allows the output bias of the op-amp to be \textbf{positive} or \textbf{negative}. As shown in Fig.~\ref{chip}, an op-amp channel consists of a differential amplification input stage, an intermediate amplification stage, and a push-pull output stage. The transfer relationship of the differential amplification input stage can be expressed as: %\Cref{opamp}
\begin{equation}\nonumber
    V_{o1}-V_{o2} = A_d(V_{i1}-V_{i2})+A_c(V_{i1}+V_{i2}) 
    \approx A_d(V_{i1}-V_{i2}) 
    \label{opamp}
\end{equation}
where $A_d$ is the differential-mode gain and $A_c$ is the common-mode gain.

\begin{figure}[t]
	\centering
	\subfigure[Input signal.]{\includegraphics[width=2cm]{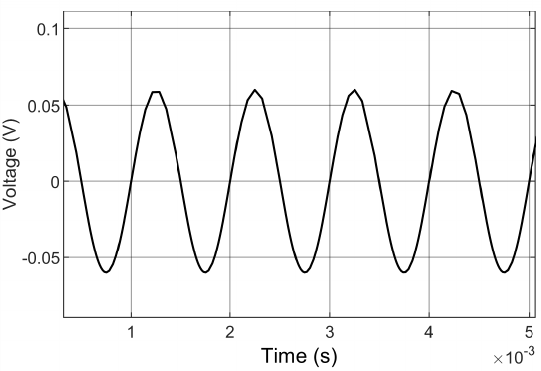}\label{input}}
      %\vspace{1mm}
     \subfigure[Output~(inject from$V_{i1}$).]{\includegraphics[width=2cm]{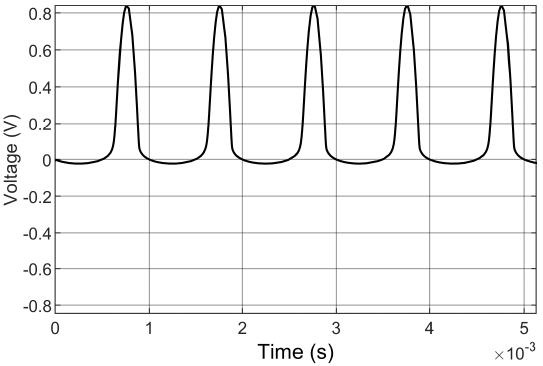}\label{o1}}
     %\vspace{1mm}
     \subfigure[Output (inject from $V_{i2}$).]{\includegraphics[width=2cm]{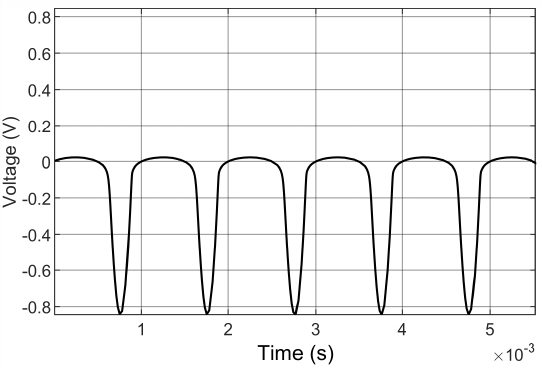}\label{o2}}
     %\vspace{1mm}
     \subfigure[Output (both inject).]{\includegraphics[width=2cm]{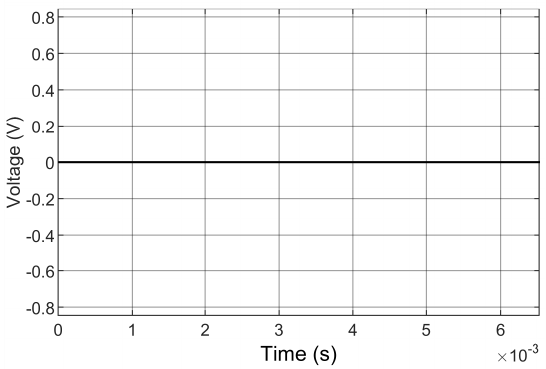}\label{NON}}
	\caption{Simulation of EMI injection on different inputs of the op-amp chip.}
\end{figure}

The asymmetric design of the input stage's wires results in different frequencies of EMI coupling. Consequently, the EMI signals coupled into $V_{i1}$ and $V_{i2}$ will differ, ultimately producing a positive or negative output. This outcome depends on whether the coupled signal is stronger at $V_{i1}$ and $V_{i2}$. To demonstrate, we build the circuit model of the OPA2171 chip in Simulink, as shown in Fig.~\ref{model} in \Cref{supplementary}. We inject the sinusoidal signal in Fig.~\ref{input} to $V_{i1}$, $V_{i2}$ or both, and we find that the output can be positive, negative, or 0 respectively, as shown in Fig.~\ref{o1}, Fig.~\ref{o2} and Fig.~\ref{NON}. Therefore, the attacker can tamper with the sensor's output to a larger or smaller value by adjusting the frequency of the EMI signal.

\textbf{\textbullet~Amplification effect.}
Amplification is the fundamental function of op-amp. Signal inputs will be amplified according to the set gain; however, EMI signals can enter into various nodes via radiative coupling. As shown in Fig.~\ref{vol}, when the EMI signal is injected into the node $b$, it can be considered that $R_1=R_2=0$. Then, according to~\Cref{A}, the gain will be abnormally large. In other words, even if injecting a millivolt signal at node $b$, it can be amplified to a few volts in process \ding{174} of Fig.~\ref{EMI_VOL}(a). 

In conclusion, electromagnetic coupling enables the injection of EMI, the nonlinear rectification converts alternative interference into positive bias, the asymmetric differential effect allows the bias to be positive or negative, and the amplification effect amplifies the injected EMI signals.

\begin{figure}[t]
	\centering
	\subfigure[Voltage sensor.]{\includegraphics[width=2.8cm]{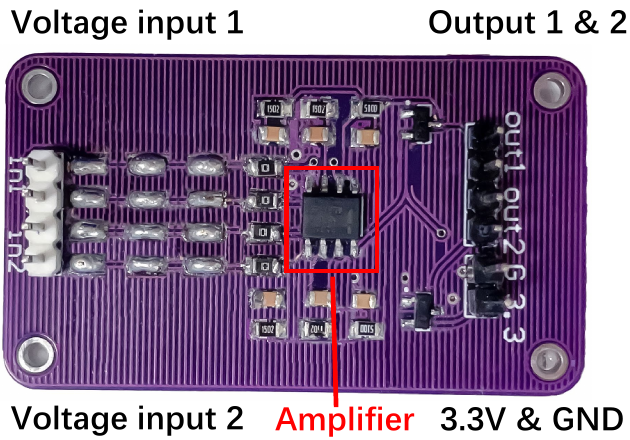}\label{s_v}}
        \hspace{10mm}
	\subfigure[Current sensor.]{\includegraphics[width=2.8cm]{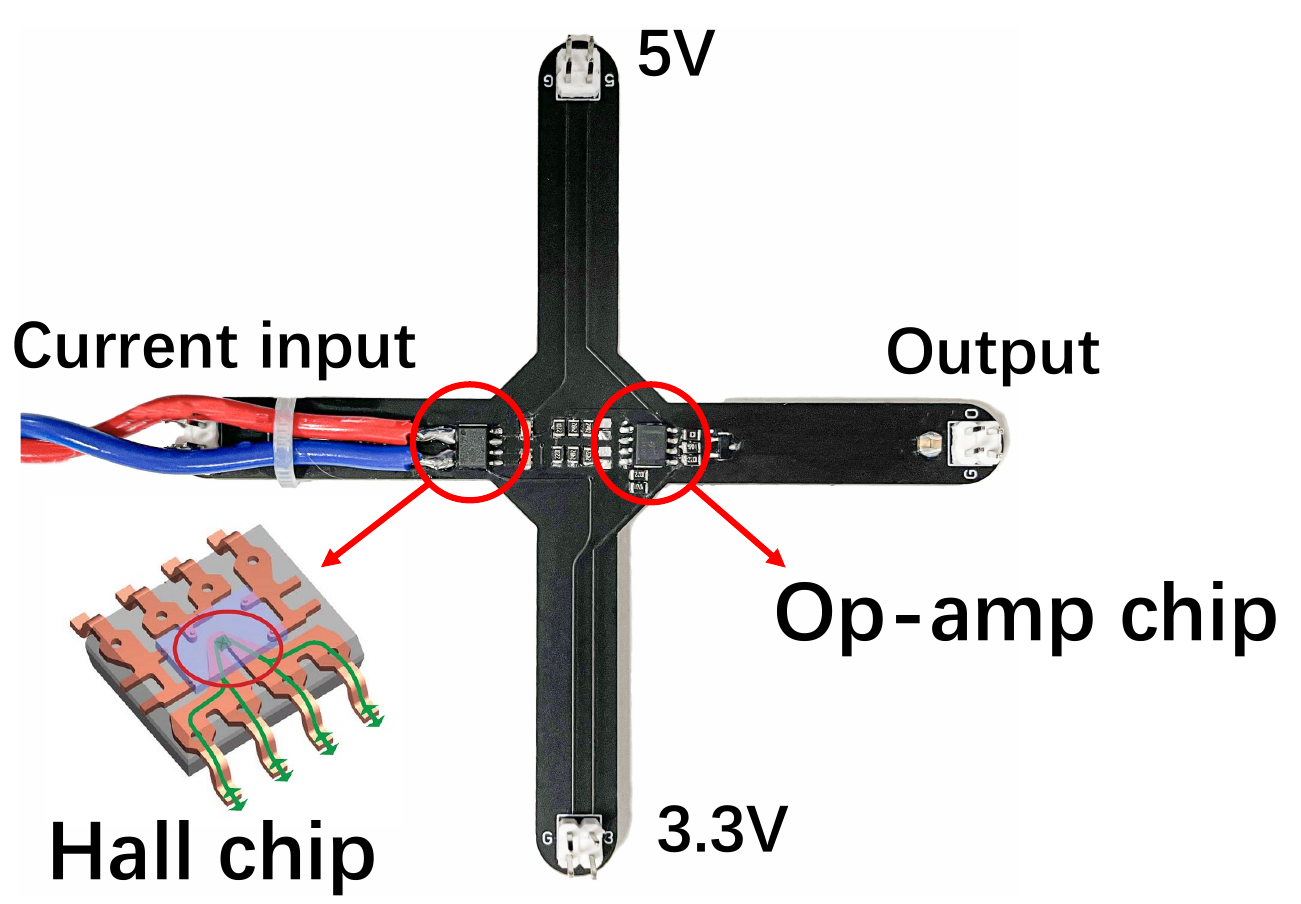}\label{s_i}}
	\caption{The voltage and current sensors' PCB we designed for the initial feasibility test.}
\end{figure}

\subsubsection{Impact of EMI on Current Sensors}

\begin{figure}[t]
%	\vspace{-1em}
	\centering{\includegraphics[width=7.6cm]{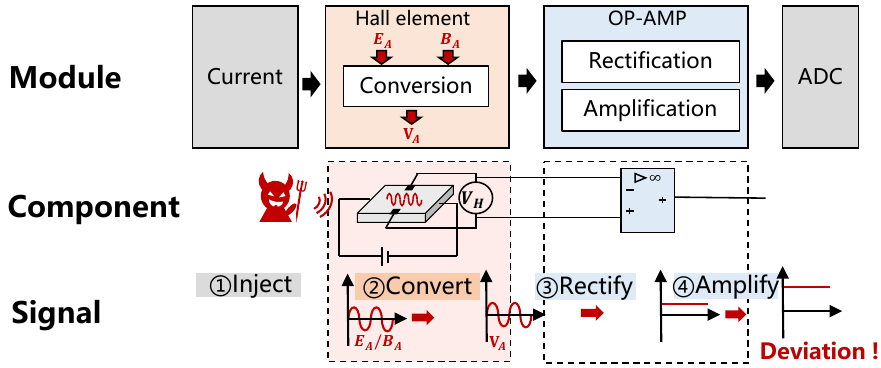}}
	\caption{The principle of EMI impact on Hall current sensors. The EMI signal is injected into the Hall chip and generates a noise $V_H$. Then the noise will be rectified, amplified by the op-amp, and result in a deviation on the output.}
	\label{cur}
	\vspace{-1em}
\end{figure}

Unlike the voltage sensor, the current sensor includes not only an op-amp circuit but also a Hall element, which may serve as a new entrance for EMI. Thus, we mainly analyze how EMI can enter the sensor circuit through the Hall chip.

We have already described that Hall current sensors measure current indirectly by measuring the magnetic field generated by the current, and the measurement relies on the balance of the Lorentz force and electric field force on the electrons, as shown in \Cref{BE}. Thus, an additional magnetic or electric field around the Hall chip will impact the current measurement, either directly or indirectly. Now we discuss them separately:

\textbf{\textbullet~Impact of magnetic field on Hall sensor.}
We assume the measured current generates a magnetic field $B$ in the Hall element. Since the output $V_{H}$ is proportional to $B$, we quantify this as~\Cref{bvd}. If EMI generates a magnetic field $B_A$ nearby, $B_A$ will be superimposed on $B$. Therefore, the output of the Hall element may be directly manipulated by the EMI signal, and this relation can described as~\Cref{vh2}, and the output $V_H$ of the Hall element will be changed by $k\cdot B_A$.
\begin{equation}
	V_{H} = k\cdot B
	\label{bvd}
\end{equation}
\begin{equation}
    V_{H}^* = k\cdot(B+B_A) = V_{H} + k\cdot B_A 
	\label{vh2}
\end{equation}

\textbf{\textbullet~Impact of electric field on Hall sensor.}
According to \Cref{BE}, we have
\begin{equation}
	V_{H} = d\cdot E
	\label{ved}
\end{equation}
If an additional electric field $E_A$ exists near the Hall chip, at this point we have
\begin{equation}
    V_{H}^* = d\cdot(E+E_A) = V_{H} + d\cdot E_A 
	\label{ved2}
\end{equation}
Thus, the output $V_H$ of the Hall chip will be changed by $d\cdot E_A$, where $d$ is the width of the electrode plate.

%%%%%%%%%%%%%%%%%%%%%%%%%%%%%%%%%%%%%%%%%%%%%%%%%%%%%%%%%%%%%%%%%%%%%%%%%%%%%%%%%%%%%%%
Then, the affected output $V_H^*$ will continue to be rectified and amplified by the op-amp and finally generate a bias on the measurement, as shown in \ding{174} and \ding{175} in Fig.~\ref{cur}.

It is worth noting that since the output $V_H$ of the Hall chip is fed into the positive input of the op-amp, %means that 
the EMI injected into the Hall chip will theoretically result in a positive bias in the current measurement. However, %since 
the EMI can also affect the op-amp of the current sensor, which will cause positive or negative bias.

\subsection{Experimental Verification}
To verify the previous analysis, we conducted feasibility tests to %assess 
explore the %impact 
capability of EMI %on 
to impact sensors of PV inverters.
\begin{figure}[t]
	\centerline{\includegraphics[width=8cm]{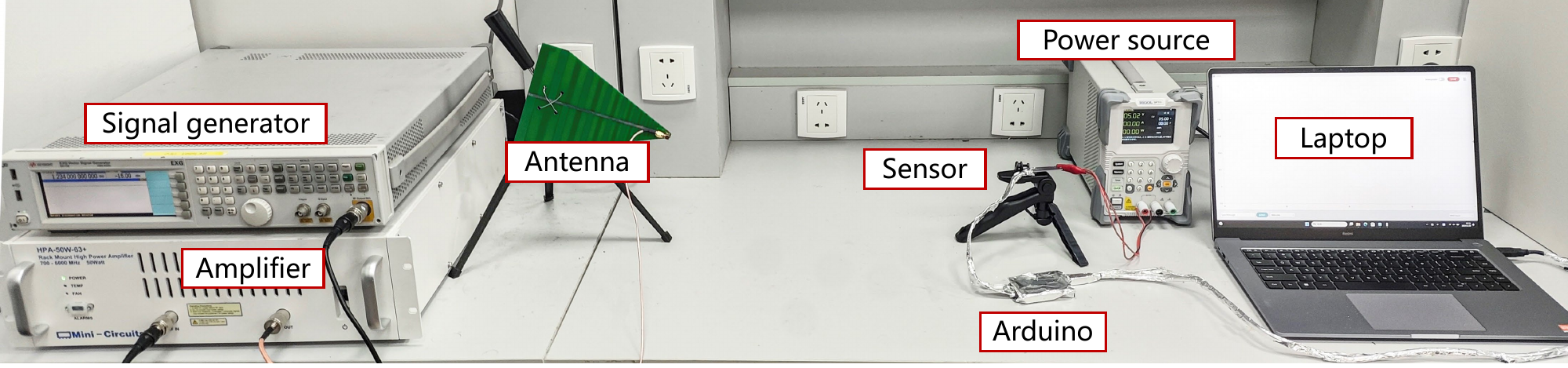}}%\vspace{-3mm}
	\caption{Setup of feasibility test on sensors.}
	\label{setup2}
\end{figure}

\subsubsection{Can EMI Impact Voltage and Current Sensors}
We conduct an EMI frequency sweep test on voltage and current sensors. The experiment setup is shown in Fig.~\ref{setup2}, and the test steps are:

\ding{172}~In the feasibility verification stage, we built the PCBs of the voltage and current sensors according to the schematic of the C2000 PV inverter from Texas Instruments (TI) that we have in hand~\cite{PCB}, as shown in Fig.~\ref{s_v} and Fig.~\ref{s_i}. 

\ding{173}~We use a DC power source RIGOL DP711~\cite{dp} to generate a $30  \, \mathrm{V}$ voltage and $0\sim 5  \, \mathrm{A}$ current to be measured. Then, we use the Arduino UNO to read the voltage every $10 \, \mathrm{ms}$ and send the data to the PC through the serial port. The Arduino is wrapped in EM shielding material to prevent EMI. All components are readily available on the market. 

\ding{174}~Subsequently, we use EXG vector signal generator~\cite{sig} to generate a $700  \, \mathrm{MHz}\sim2.5  \, \mathrm{GHz}$ signal, use amplifier HPA-50W-63+~\cite{amplifier} to amplify it to $10  \, \mathrm{W}$, and emit it with a 5G directional antenna~\cite{antenna} with $+14  \, \mathrm{dBi}$ at a distance of $50  \, \mathrm{cm}$.   
\begin{figure}[t]
	\centering
	\subfigure[Voltage sensor.]{\includegraphics[width=4.1cm]{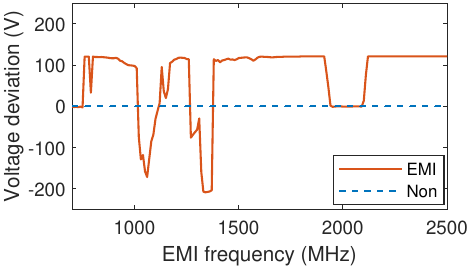}\label{av}}
	\subfigure[Current sensor.]{\includegraphics[width=4.1cm]{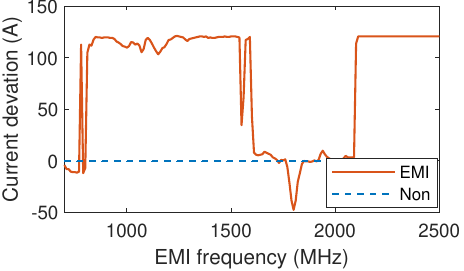}\label{ai}}
	\caption{The result of the EMI frequency test on the voltage and current sensors. The EMI power and distance are set to $10 \, \mathrm{W}$ and $50 \, \mathrm{cm}$.}
	\label{avi}
	%\vspace{-1em}
\end{figure}

We record the deviation of the measurements in Fig.~\ref{avi}. For the voltage sensor, the measured voltage can be decreased by $200  \, \mathrm{V}$ and increased by $120  \, \mathrm{V}$ at most. For the current sensor, the measured current can be increased by up to $320  \, \mathrm{A}$ and decreased by up to $30  \, \mathrm{A}$. The result demonstrates that EMI can effectively affect the voltage and current sensor's outputs. Notably, in the test of the Hall current sensor, the deviation of the measurement is predominantly positive. This verifies our previous analysis of the impact of EMI on current sensors. 

To further verify that the EMI can impact the Hall chip directly, we conducted a small test: We measured the output $V_H$ of the Hall chip using RF wines to avoid wire coupling, with the sample rate of $10  \, \mathrm{GHz}$, and compare the effect of EMI on $V_H$. The result is shown in Fig.~\ref{con} in \Cref{supplementary}. It shows that EMI can directly impact the Hall chip by inducing a $0.2  \, \mathrm{V}$ bias and a $0.5  \, \mathrm{V}$ oscillation on the output $V_H$ of the Hall element. 

\subsubsection{Whether the Impact is Controllable}

To explore the EMI manipulation capability on sensors, we tested two kinds of EMI signal modulation methods: 

\ding{172}~Frequency modulation~(FM). Fig.~\ref{av} and Fig.~\ref{ai} reveal that sensors have different ``sensitivity'' to EMI signals of various frequencies. It appears that adjusting the signal frequency may manipulate the target sensor's output. However, we can also find that the sensor's output varies significantly as the frequency changes. Therefore, achieving precise control of sensor values with FM proves challenging.

\ding{173}~Amplitude modulation~(AM). Another signal modulation method is AM, as described in~\Cref{sam}, where $s_m(t)$ is the modulation signal, $A_c$ and $f_c$ are the amplitude and frequency of the carrier signal $s_c(t)$.
\begin{equation}
	s_{AM}(t) = A_c[1+s_m(t)]cos2\pi f_ct 
	\label{sam}
\end{equation}

%根据，，，传感器输出的偏移量和攻击信号的幅度成正比.然后将s设置成我们想让传感器变化的曲线，也就是s2的包络线。
Since the offset of the sensor's output is proportional to the amplitude of the EMI signal, we first select a carrier signal $s_c(t)$ that can impact the sensor's output, then set $s_m(t)$ to the ``desired'' curve, which is also the envelope of $s_{AM}(t)$.  

\begin{figure}[t]
	\centerline{\includegraphics[width=8.9cm]{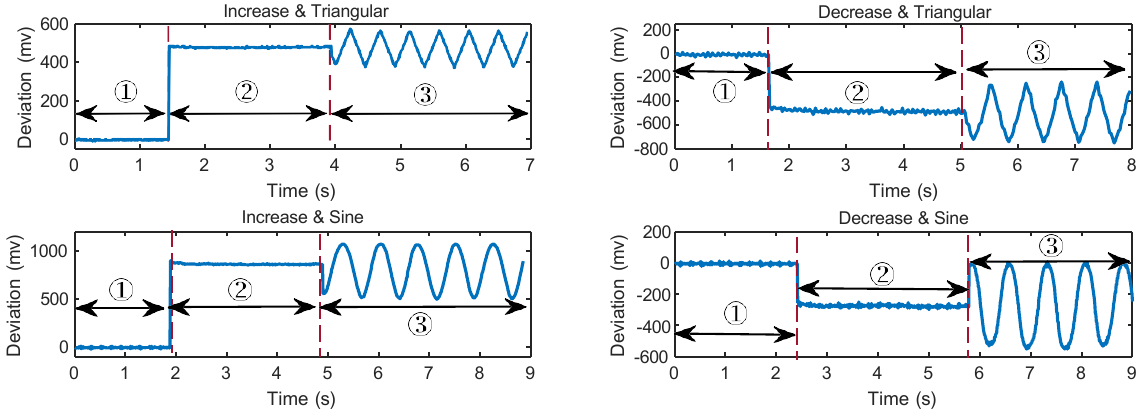}}%\vspace{-3mm}
	\caption{The experiment result of manipulation with a single-frequency signal and an AM signal on the sensor. \ding{172}:Without EMI; \ding{173}:Single-frequency EMI; \ding{174}:AM-modulated EMI.}
	\label{siheyi}%\vspace{-5mm}
\end{figure}

In this scenario, assuming that one wants the measured voltage to first increase or decrease and then change as the triangular or sine wave, we conducted an experiment using AM. The result is %entirely ``satisfactory''
highly ``favorable'' for an adversary, as depicted in Fig.~\ref{siheyi}. Although the real voltage or current remains constant, the measured values change precisely by the $s_m(t)$, such as triangular and sine waves. 

\subsubsection{Verification of the Universality and Extensibility}
Commercial PV inverters usually contain multiple types of sensors. To analyze the universality of the threat, we propose two questions: \ding{172}~What is the impact of EMI on different Hall sensors? \ding{173}~If there are multiple sensors, can EMI only impact a single target sensor or control multiple target sensors simultaneously?

\textbf{Universality.}~To answer the first question, we evaluate the impact of EMI on 7 different Hall sensors, including 4 analog sensors and 3 digital sensors. Hall digital sensors include a speed sensor, a north pole sensor, and a water flow sensor. The result is presented in \Cref{sensors}. We can find that both wired and wireless Hall current sensors are susceptible to EMI, and wireless Hall current sensors exhibit a higher degree of susceptibility. Hall sensors with digital outputs, like speed sensors, may experience bit-flipping under EMI.

\textbf{Extensibility.~}Since EMI signals of different frequencies can be injected into different nodes of the victim circuit, we can establish a frequency sweep model for each sensor and implement the following: ~\ding{172} ``one-to-one'' manipulation: select a frequency that exclusively affects the target sensor without impacting others; ~\ding{173} ``many-to-many'' manipulation: when manipulating several sensors simultaneously, owing to the superposition of EMI signals, we can employ different channels to emit EMI signals of various frequencies. \textit{This feature also highlights one of the advantages of EMI over constant magnetic field attacks in Hallspoofing~\cite{barua2020hall}: higher extensibility in signal design through signal multiplexing.}

\begin{table*}[pt]
	%\vspace{-1em}
	\caption{Result of EMI impact on 7 Hall sensors} 
	\begin{center}
            \setlength\tabcolsep{8pt} %调整列间距，即表格宽度
		\begin{threeparttable}
                \renewcommand{\arraystretch}{0.9}
			\begin{tabular}{c|c|c|c|c|c|c|c|c|c|c}
				\toprule[2pt]
				\multirow{2}{*}{\makecell{Sensor \\ type}} & \multirow{2}{*}{\makecell{Sensor \\ model}} & \multirow{2}{*}{\makecell{Output \\type}} & \multirow{2}{*}{\makecell{Measure-\\ment \\ span}} & \multicolumn{2}{c|}{Test parameters} & \multicolumn{5}{c}{Output}\\
				\cline{5-11}
				
				&                        &                              &                                   & \makecell{Freq.(MHz)\tnote{3} \\ (Pos. /Neg.)\tnote{3}} & Pow.(W)\tnote{3} & \makecell{Original \\value} & \makecell{Pos. \\ dev.\tnote{1,3}} & \makecell{Pos. \\ dev. rate}    & \makecell{Neg. \\ dev.\tnote{1,3}} & \makecell{Neg. \\ dev. rate} \\
				\hline
				\makecell{Current} & \makecell{WCS1800~(Wire)} & Analog & 0$\sim$30A & 685/1030 & 10 & 5~A & 15.7~A & +214.00\% & -6.1~A & -222.00\% \\
				\hline
				\makecell{Current } & \makecell{WCS1800~(Wireless)}      & Analog                       & 0$\sim$35~A                             & 1000/876                                        & 10      & 5~A        & 31.5~A             & +530.00\% & -7.6~A             & -252.00\% \\
				\hline
				\makecell{Current }  & ACS712~(20~A)                 & Analog                       & 0$\sim$20~A                             & 779/1223                                        & 10      & 5~A        & 13.2~A            & +164.00\% & -13.2~A            & -364.00\% \\
				\hline
				\makecell{Current }  & ACS712~(5~A)                & Analog                       & 0$\sim$5~A                              & 627/1212                                        & 10      & 2.5~A      & 5.1~A              & +104.00\% & -7.75~A            & -410.00\% \\
				\hline
				\makecell{Speed }  & 3144                   & Digital                      & 0/1                               & 677                                             & 10      & 0/1        & bit-flap\tnote{2}                & +100.00\% & bit-flap           & -100.00\%    \\
				\hline
				\makecell{North pole }            & 3144                   & Digital                      & 0/1                               & 724                                             & 10      & 0/1        & bit-flap                & +100.00\% & bit-flap           & -100.00\%    \\
				\hline
				\makecell{Water flow }            & YF-S401                & Digital                      & 0/1                               & 1322                                            & 10      & 0/1        & bit-flap                & +100.00\% & bit-flap           & -100.00\%  \\ 	
				\bottomrule[2pt]
			\end{tabular}
			\begin{tablenotes}
				\item[1] For each Hall current sensor, we repeat each experiment 10 times and calculate the average deviation.
				\item[2] For each Hall digital sensor, we only record whether the output experiences a bit-flip.
				\item[3] ``Freq.'' means frequency; ``pos.'' and ``neg.'' means positive and negative deviation; ``dev.'' means deviation.
			\end{tablenotes}
		\end{threeparttable}
		\label{sensors}	 
	\end{center}
	\vspace{-1em}
\end{table*}

% !TEX root = ../Theremin.tex
\section{Understanding the Impact of Sensor Spoofing on PV Inverters}
\label{sc5}
Here, we analyze how the spoofing of sensors affects the operation of PV inverters. We build the PV inverter circuit model and implement the control algorithms outlined in~Section \ref{sc2} using Simulink. 

\subsection{Impact of DC Bus Voltage Sensor}
Deceiving the DC bus sensor will directly affect the DC bus voltage control loop. The function of the voltage control loop is to maintain the DC bus voltage $V_{dc}$ as its reference value $V_{dcref}$ set by the manufacturer. When an EMI signal introduces a deviation of $V_a$ on the measured bus voltage, it will lead to \Cref{vdc}:
\begin{equation}
	V_{dc}^* := V_{dc}+ V_a \, 
	\label{vdc}
\end{equation}
Then the controller will adjust $V_{dc}^*$ to be equal to $V_{dcref}$, and the real DC bus voltage will become $V_{dcref}- V_a$ under control. This will cause the following damages. %to the PV inverter:

\subsubsection{Breakdown of DC Bus Capacitor~($V_a < 0$)} If the EMI signal introduces a negative $V_a$ to the measured $V_{dc}$, the real DC bus voltage will increase and the aging of the DC bus capacitor $C_{dc}$ will accelerate. The capacitor will break down when the voltage exceeds the rated voltage of the $C_{dc}$. 
While the inverter incorporates over-voltage and under-voltage protection mechanisms, the vulnerability could persist, potentially leading to physical damage. This risk emerges when the adversary intentionally avoids injecting $V_a$ with a substantial magnitude in a single instance. This is attributed to continuously manipulating sensor values to appear within their normal range while the real DC bus voltage is spoofed. For the adversary, he may want to ensure that, during the injection of the EMI signal, the sensor value does not trigger the under-voltage protection mechanism, allowing the EMI to circumvent the protective measures. Afterward, the inverter loses its ability to operate correctly due to the deficiency in the $C_{dc}$'s capacity to balance the input and output power. 

\begin{figure}[tb]
%\vspace{-1em}
\centering
\subfigure[$V_a < 0$.]{\includegraphics[width=4.1cm]{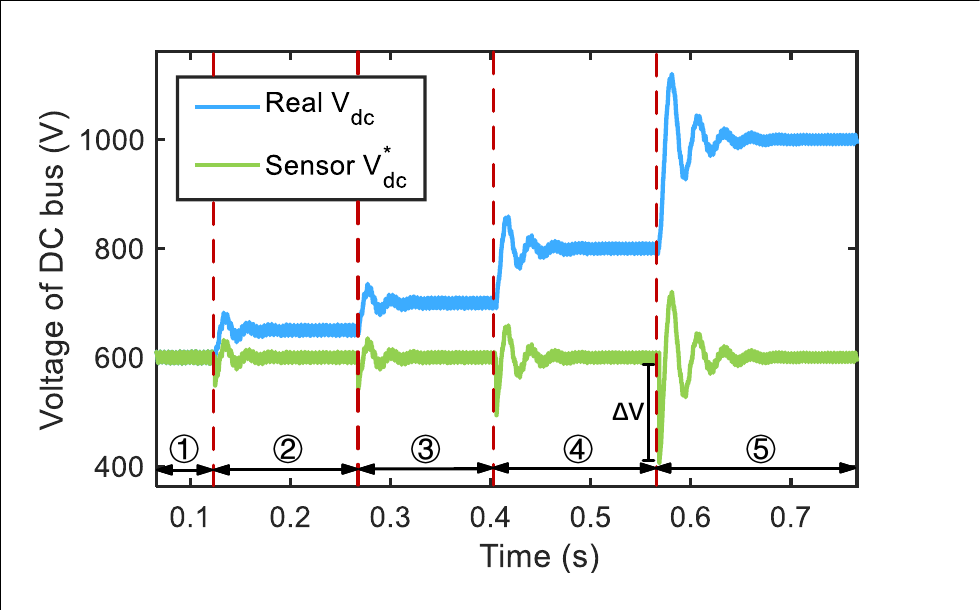}\label{figVdc_rd}}
\hspace{2mm}
\subfigure[$V_a > 0$.]{\includegraphics[width=4.1cm]{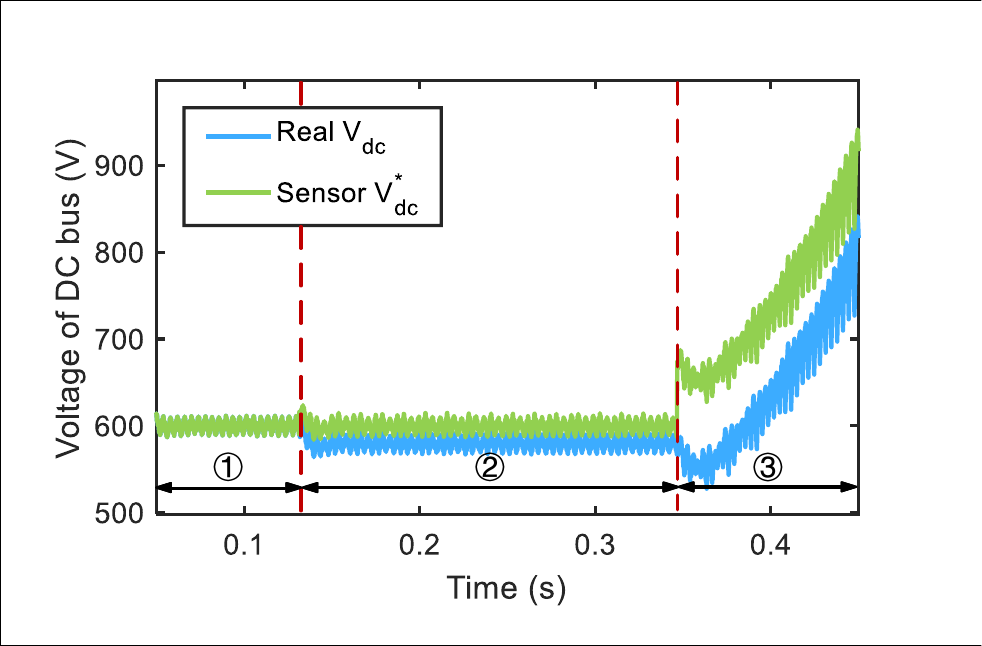}\label{figVdc_rs}}
\caption{The simulation of the DC bus voltage manipulation. We add a fake $V_a$ on the measured DC bus voltage and record the real DC bus voltage under control. For $V_a < 0$, \ding{172}: $V_a = 0V$, \ding{173}: $V_a = -50V$, \ding{174}: $V_a = -100V$, \ding {175}$V_a = -200V$, \ding {176}$V_a = -300V$;  for $V_a > 0$, \ding{172}: $V_a = 0V$, \ding{173}: $V_a = 20V$, \ding{174}: $V_a = 100V$.}
%\vspace{-1em}
\end{figure}

The simulation results are given in Fig.~\ref{figVdc_rd}. It can be observed that the real DC bus voltage is increased by $50  \, \mathrm{V}$, $100  \, \mathrm{V}$, $200  \, \mathrm{V}$ and $300  \, \mathrm{V}$ after sensor manipulation. Looking at Fig.~\ref{figVdc_rd} for the case of $V_a = -300  \, \mathrm{V}$, the transient voltage offset $\Delta V$ will trigger the protection instantly and shut down the inverter. %will be shut down.

\subsubsection{DC Bus Under-voltage~($V_a > 0$)} Similarly, an adversary can decrease the real DC bus voltage by injecting a positive $V_a$ into the voltage measurement. If the real DC bus voltage drops below the lower threshold, the output AC voltage will be lower than the grid voltage. In that case, the current will be reversed, and the power will flow back from the grid to the inverter, and the protection mechanisms will be triggered to shut down the inverter. This process is shown in Fig.~\ref{figVdc_rs} when $V_{a}=100  \, \mathrm{V}$. 

Hence, in summary, the impact of sensor spoofing on the DC bus voltage can be articulated as follows: 

\textbf{Impact 1: \dos.} The \dos stops the PV inverter's normal operation. The key of \dos is to trigger the self-protection mechanism of PV inverters. As previously analyzed, there exist two methods to induce \dos. Here, we illustrate the process by taking the example of injecting a positive deviation ($V_a > 0$) on the DC bus voltage sensor. To achieve this objective, the adversary could design the EMI by the following steps: 

To begin, it is imperative to carefully select the frequency $f_{c+}$ of the EMI signal through preliminary frequency testing. This choice can potentially augment the measured $V_{dc}$. Given that PV inverters of similar application levels, such as residential PV inverters ranging from $1$ kW to $60$ kW, typically share similar PCB dimensions, the frequencies susceptible to EMI do not show substantial variations. Drawing from our empirical observations, $f_{c+}$ commonly falls within the range of $700 \, \mathrm{MHz}$ to $1500 \, \mathrm{MHz}$. Subsequently, as the adversary approaches the PV inverter, it becomes necessary to transmit the EMI signal at the designated frequency $f_{c+}$ for a brief duration, typically spanning a few seconds.

\textbf{Impact 2: \damage.} \damage can potentially result in the permanent breakdown of the DC bus capacitor and inflict harm upon the PV inverter. To effectuate \damage, an adversary must elevate the real $V_{dc}$ by introducing a negative %voltage (
$V_a$ into the measured $V_{dc}$ while circumventing the activation of the %inherent 
self-protection mechanism. 

\begin{figure}[t]
	\centering
	\subfigure[EMI signal for \dos on AC side.]{\includegraphics[width=4.2cm]{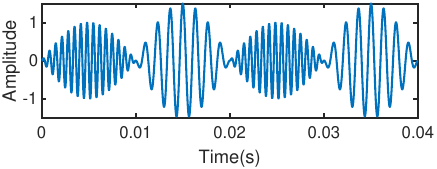}\label{am_sin}}
        \hspace{1mm}
	\subfigure[EMI signal for %principle of 
	\damage.]{\includegraphics[width=4.2cm]{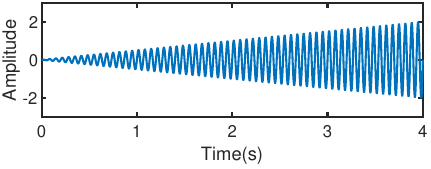}\label{am_step}}
	\caption{Design of EMI signals $s(t)$ of \dos and \damage.}
\end{figure}

First, the adversary needs to find the frequency $f_{c-}$ that can efficiently decrease the measurement of $V_{dc}$ and generate the carrier signal $s_c(t)$. Since the victim system takes time to reach the stability of $V_{dc}$ after each manipulation, the adversary can design $s_m(t)$ as, \Cref{mt}, where $k$ and $s_0$ are the scale factor and initial value of $s_m(t)$. Generally, the smaller $k$ is, the easier it is to avoid triggering the self-protection mechanism, but it takes a longer time. Finally, the adversary obtains $s(t)$ by AM, as shown in Fig.~\ref{am_step}.  
\begin{equation}
	s_m(t) = kt + s_0,~k>0,~ s_0\ge0 
	\label{mt}
\end{equation}

To avoid triggering the protection mechanism, for the TI C2000 PV inverter~\cite{QSG}, %that we have in hand, 
the target $V_{dc}$ is $385  \, \mathrm{V}$, and the safety range is $220  \, \mathrm{V} \sim 395  \, \mathrm{V}$. %which means 
It indicates that the adversary needs to allow time for the controller to adjust  $V_{dc}$ within %the range of  $220  \, \mathrm{V} \sim 395  \, \mathrm{V}$, 
this range after each manipulation.

\subsection{Impact of Grid Voltage and Current Sensors}
\label{sc422}

The measured grid voltage and current serve as feedback for the current control loop. Manipulations on these sensors have different effects on single-phase and three-phase PV inverters. The three-phase inverter supplies a three-phase AC power output that the phases are $120  \, \mathrm{^{\circ}}$ between each other, commonly used in industrial and commercial settings. The single-phase inverter outputs one-phase AC power, typically employed in residential PV generations.

\subsubsection{Single-phase PV Inverter} 
We take the manipulation of grid current as an instance. If the injected deviation $I_{a}$ is constant, there will be a ``transient effect'' on the real grid current. This is similar to the case in which the inverter suffers from sudden grid current changes while the control loops manage to restore the current. To illustrate, let $I_a$ be constant and positive, then the controller will decrease the current, and the inverter's output power will decrease. However, when the output power becomes less than the input power, the DC bus capacitor will charge, leading to $V_{dc} > V_{dcref}$, and the current reference will increase. In this regard, the reference will rise again to catch up with the manipulated current. 

\begin{figure}[tb]
\centering
\subfigure[Single-phase PV inverter. \ding{172}:$I_a= 0A$, \ding{173}:$I_a= 50\sin \omega t A$, \ding{174}:$I_a= 200\sin \omega t A$.]{\includegraphics[width=4.1cm]{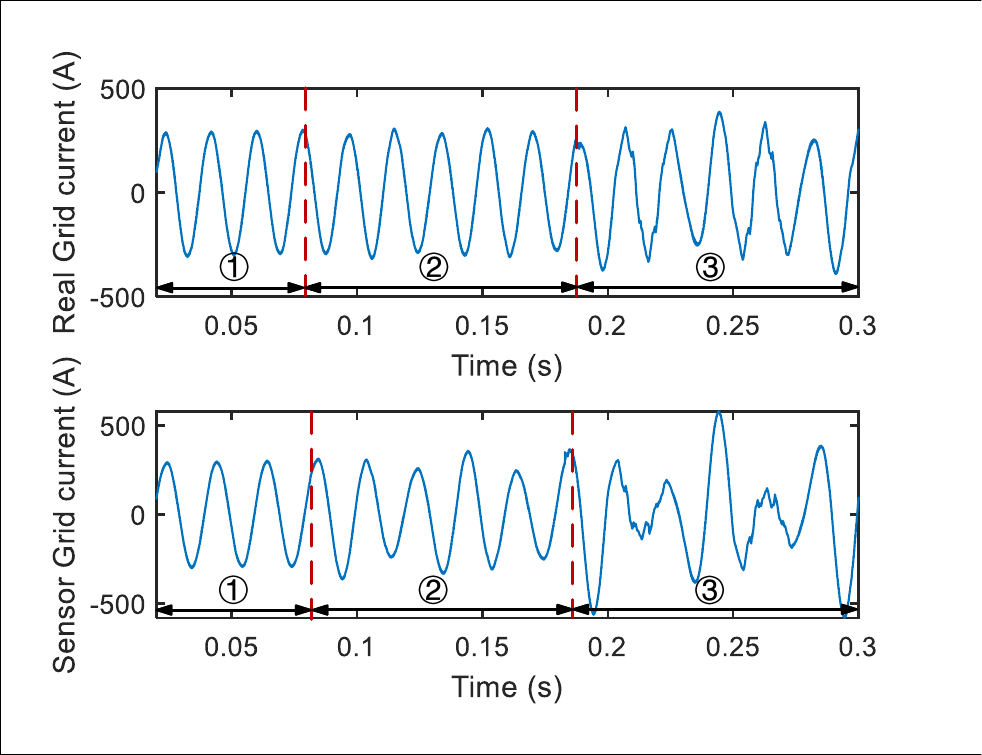}\label{Girdsigphase_Il}}
\hspace{2mm}
\subfigure[Three-phase PV inverter. \ding{172}:$I_a= 0A$, \ding{173}:$I_a= 50A$, \ding{174}:$I_a= 200A$.]{\includegraphics[width=4.1cm]{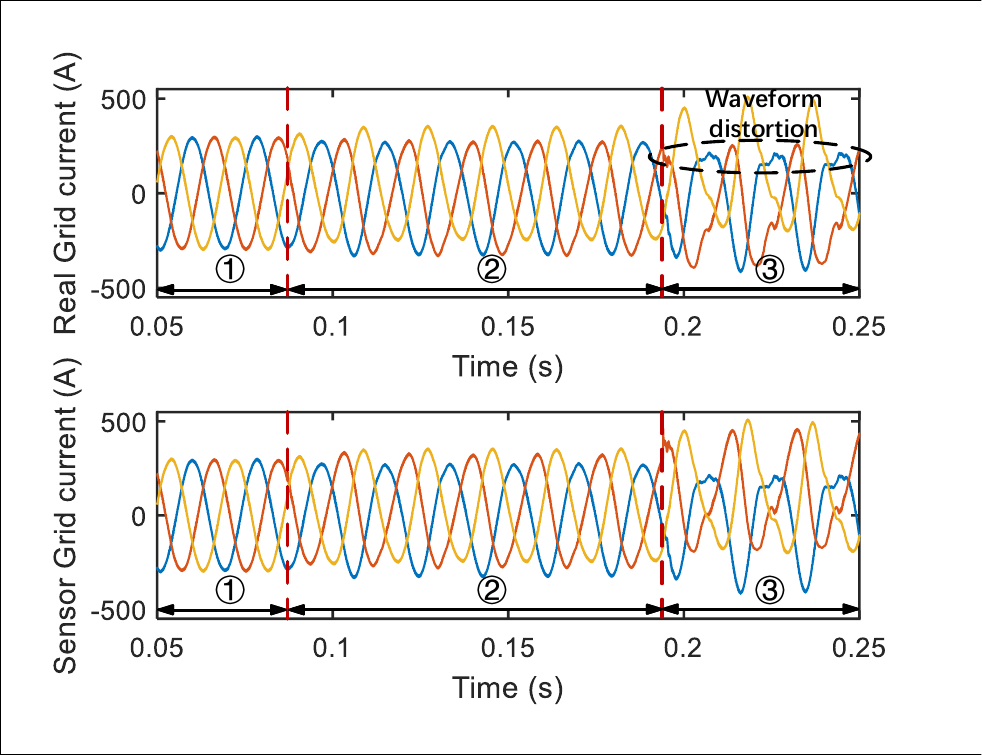}\label{Girdsig_Iabc}}
\caption{The simulations of grid current sensors spoofing. It gives the simulated waveform of the real current value and the sensor output value when the single-phase and three-phase grid current measurement is manipulated.}
%\vspace{-1em}
\end{figure}

To note, if the injected deviation $I_{a}$ is time-varying, like a 
sinusoidal signal, the PV inverter will not enter into a steady state. The simulation result is shown in Fig.~\ref{Girdsigphase_Il}. The larger the magnitude of the injected deviation $I_a$, the higher the degree of oscillation in the grid current. When the oscillation reaches a certain level, the grid current and voltage will exceed the threshold and trigger the protection mechanism, and the inverter will shut down.

\subsubsection{Three-phase PV Inverter}  As mentioned in the background, the three-phase voltage and current output of the PV inverter need to be transformed into the coordinate system through the Clark transformation and Park transformation before entering the control loop. In fact, due to this coordinate system transformation, a constant injected deviation into the three-phase voltage and current measurements could not affect the inverter's output. This is because it will be filtered out by the Clark transformation matrix. \textit{Thus the Hallspoofing attacks in \cite{barua2020hall} may fail in such scenario.} 

Therefore, the impact of grid voltage and current sensor manipulation in the three-phase PV inverter will only manifest when the injections are ``unequal''. As illustrated in Fig.~\ref{Girdsig_Iabc}, compared to the single-phase inverter that needs to inject a time-varying $I_{a}$, the three-phase inverter only needs to inject a constant $I_{a}$ into one phase but not other phases to achieve a similar impact~(inverter shutting down). The coordinate system is time-varying, making the component on each axis of the time-invariant signal also time-varying. For simulations of measurement manipulation on grid voltages, we refer to \Cref{Gridvoltagetampering}. We now summarize the impact of grid voltage and current sensor spoofing on PV inverters:

\textbf{Impact: \dos.} For \dos impact on the grid AC side, the primary adversarial strategy involves inducing oscillations in the AC voltage or current. Taking the AC current as an example, the adversary needs to inject a time-varying signal $I_a(t)$ on the measured AC current. We select $I_a(t)$ as a sine wave with the same frequency as the AC, which is not the only option.
\begin{equation}
	I_a(t) = A_{a} \cdot sin(2\pi f_{AC}t) 
\end{equation}
where $f_{AC}$ is the AC frequency, and $A_a$ is the amplitude of $I_a(t)$. Since the grid imposes strict limitations on input voltage and current, an $I_a(t)$ with a few amps is enough to achieve the impact of \dos. 

First, the adversary needs to find the frequency $f_{c+}$ and $f_{c-}$ that can increase and decrease the measured AC current. %$I_{AC}$. 
Then he may design the modulation signal $s_m(t)$ as:
\begin{equation}
	s_m(t) = sin(2\pi f_{AC}t) 
\end{equation}

Finally, get the EMI signal $s(t)$, as shown in Fig.~\ref{am_sin}, 
%\newpage

\begin{equation}
		\begin{split}
			s(t) = \left\{
			\begin{array}{lr}
				A_+(1+s_m(t))cos2\pi f_{c+},   s_m(t) > 0,  \\
				A_-(1+s_m(t))cos2\pi f_{c-},   s_m(t) \le 0 
			\end{array}
			\right.
		\end{split}
\end{equation}

The adversary only needs to continuously transmit the signal for a few seconds when passing by the target inverter.

\subsection{Impact of PV Voltage and Current Sensors}
\label{gridtamper}
The PV voltage and current sensors are used for the MPPT algorithm and the DC-DC stage. Since the MPPT algorithm regulates the input power of the inverter by controlling the input voltage, manipulating $V_{pv}$ and $I_{pv}$ can impact the output power of the PV inverter. 

Injecting a constant offset $\Delta V$ on the PV voltage sensor or $\Delta I$ on the PV current sensor~(Fig.~\ref{PVinject1} in~\Cref{grid}) only shifts the V-I curve without changing its ``shape''. Thus, the MPPT algorithm will still find the correct MPP with false measured $V_{pv}$ or $I_{pv}$ by the P\&O algorithm.

However, if the adversary can design a fake V-I curve with a different shape from the original one, the MPPT algorithm will be misled into finding the fake MPP, resulting in decreased power. To inject a fake V-I curve, the adversary needs to make the spoofed points~($V_{pv}$, $I_{pv}$) move on a fixed but false curve by manipulating the measured $I_{pv}$ or $V_{pv}$, as shown in Fig.~\ref{PVinject3} in~\Cref{grid}. 
We will specify this method in the following. 

\textbf{Impact: \damp.} \damp will adversely impact the efficiency and reduce the output power of PV inverters.
The primary objective of the \damp is to deceive the MPPT algorithm, preventing it from accurately identifying the MPP. Two distinct EMI design strategies for achieving this objective exist, categorized as ``spoofing'' and ``interference''. The ``spoofing''-based method quantitatively diminishes the power output of the target PV inverter but necessitates the utilization of feedback information, namely $V_{pv}$ and $I_{pv}$ values from the internal sensors of the PV inverter. Conversely, the ``interference''-based method can relatively reduce the power of the PV inverter without requiring any feedback information. 
 %in the following steps:

For the \damp based on ``interference'': Since the MPPT finds the MPP by P\&Q method that relies on stable $V_{pv}$ and $I_{pv}$, the adversary could tamper with $V_{pv}$ or $I_{pv}$ to interfere the MPPT. The EMI threat can be designed akin to the \dos scenario to disrupt the measurement of $V_{pv}$ or $I_{pv}$, thereby impeding the MPPT algorithm from achieving maximum power. According to our experiment on the TI C2000 PV inverter, the injected $V_{a}$ should be between $-5  \, \mathrm{V}$ and $+5  \, \mathrm{V}$ to avoid triggering \dos impact instead; this threshold can be obtained by pre-test. 

For the \damp based on ``spoofing'': First, an adversary needs to acquire the V-I or V-P characteristic curve $f(V, I)$ from the user's manual of the PV panel. Without this information, the attacker needs to buy the same model of PV panel and measure the voltage and current under different insolation and temperatures to plot the V-I curve. Then, the adversary needs to design a fake V-I curve $f^*(V, I)$, and the MPPT algorithm will reach the false MPP on the fake V-I curve. %$f^*(V, I)$. 
The real power will be reduced by $\Delta P$, as shown in Fig.~\ref{WeakenAttack1}. There are %different 
diverse fake V-P curves, as indicated in Fig.~\ref{WeakenAttack2}. The challenge lies in making the point~($V_{pv}$, $I_{pv}$) move on the fake V-I curve $f^*(V, I)$ at any time. To achieve this, one feasible solution is to adjust $I_{pv}$ according to $V_{pv}$, as illustrated in Fig.~\ref{WeakenAttack3}, and the steps can be summarized as follows: 

\begin{figure}[t]
%	\vspace{-1em}
	\centering
	\subfigure[V-P curve.]{\includegraphics[width=2.7cm]{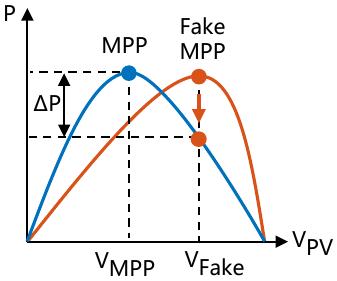}\label{WeakenAttack1}}
        \hspace{2mm}
	\subfigure[Diversity of the curve.]{\includegraphics[width=2.7cm]{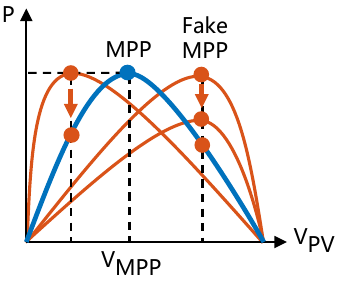}\label{WeakenAttack2}}
        \hspace{1mm}
	\subfigure[V-I curve.]{\includegraphics[width=2.7cm]{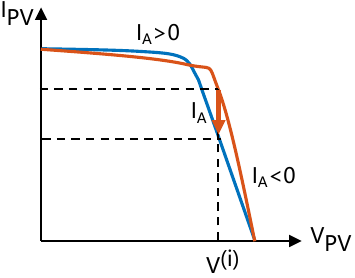}\label{WeakenAttack3}}
	\caption{False V-I curve design of \damp based on the ``spoofing'' method.}
%	\vspace{-1em}
\end{figure}

First, the adversary needs to design a fake V-I curve $f^*(V, I)$. When approaching the victim PV inverter, the adversary reads the $V_{pv}$ and $I_{pv}$ via the pre-installed sensors in the inverter as the feedback information and adjusts $I_{pv}$ to the fake V-I curve by adjusting the EMI power. The pseudo-code is illustrated in \Cref{alg}.

\begin{algorithm}
	\renewcommand{\algorithmicrequire}{\textbf{Input:}}
	\renewcommand{\algorithmicensure}{\textbf{Output:}} 
	\caption{\damp based on ``spoofing''} 
	\label{alg} 
	\begin{algorithmic}
		\REQUIRE Measured PV voltage $V_{pv}$, measured PV current $I_{pv}$
		\ENSURE  EMI power $P$
		\WHILE{1}
		\STATE Read $V_{pv}$
		\STATE Read $I_{pv}$
		\STATE Get expected $I^*_{pv}$ according to $V_{pv}$ and fake V-I curve
		\IF{$I_{pv}$ < $I^*_{pv}$}
		\STATE Adjust EMI power $P_A$ to increase $I_{pv}$
		\ELSE
		\STATE Adjust EMI power $P_A$ to decrease $I_{pv}$
		\ENDIF
		\STATE Wait inverter to update $V_{pv}$ and $I_{pv}$
		\ENDWHILE 
	\end{algorithmic} 
\end{algorithm}
%

% !TEX root = ../Theremin.tex
\section{Threat Evaluation}

In this section, we first evaluate the threat of \alias on PV inverters, and then test on a rural-scale microgrid operated in a real world to explore the impact of \alias on the grid. To our knowledge, this is the first work validating EMI threat on the real-world microgrid. \textit{To ensure the safety and legality of the research, we conducted all indoor experiments in an electromagnetic shielding room, and we contacted the manufacturer and local distribution grid operator about the testing details to avoid ethical problems.}

\subsection{Evaluation on PV Inverters}
\subsubsection{Experiment Setup}
\begin{figure}[t]
	\centerline{\includegraphics[width=8.4cm]{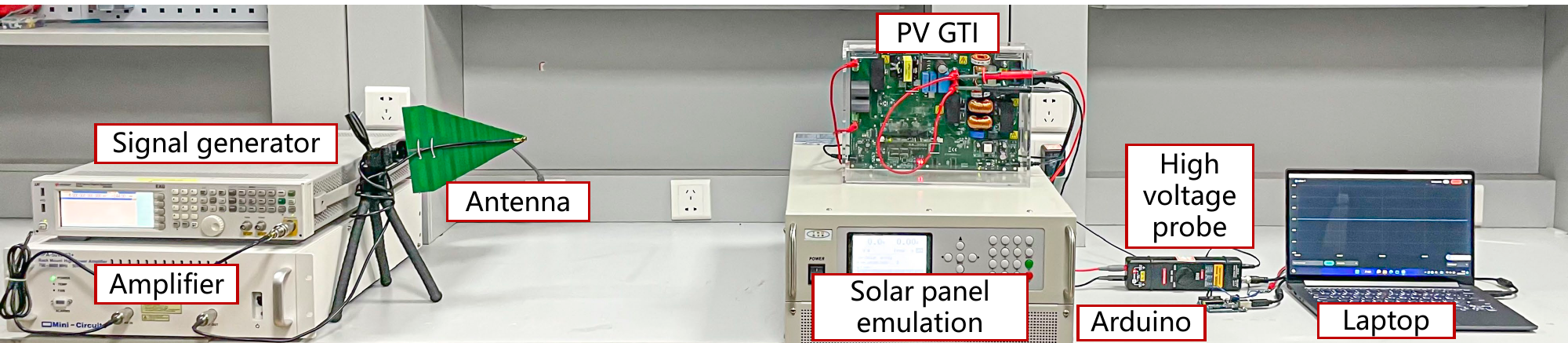}}%\vspace{-3mm}
	\caption{Experiment setup of evaluation on PV inverters.}
	\label{setup1}
	\vspace{-1em}
\end{figure}

As shown in Fig.~\ref{setup1}, the experimental setup comprises victim and adversary devices. The victim devices are off-the-shelf %commercial 
PV inverters, and adversary devices are used to emit EMI signals. 

\textbf{Victim Devices.} 
To verify the universality of the threat, we select a TI C2000 inverter development kit~\cite{QSG} and 5 commercial PV inverters ~\cite{Ginlong}~\cite{Kstar}~\cite{huawei}~\cite{Goodwe50} from well-selling manufacturers, as shown in Fig.~\ref{zhaopian} in \Cref{supplementary}. The inverters~\cite{QSG}~\cite{Ginlong}~\cite{Kstar}~\cite{huawei} are tested under laboratory conditions, and 2 models of inverters designed by GoodWe~\cite{Goodwe50} are tested in a real-world microgrid.

Compared with commercial inverters, the TI inverter development kit has the following features:~\ding{172}~lower power and higher safety;~\ding{173}~most of the process variables can be read from the upper computer;~\ding{174}~open-source control programs. %software. 
In comparison, commercial PV inverters \ding{172}~have better EMC countermeasures (such as special enclosures and internal filtering circuits); \ding{173}~operate at higher power levels~(several kWs), posing risks for conducting \damage experiments; thus we evaluate all three impacts of \alias on the C2000 solar micro inverter and evaluate \dos and \damp on 5 commercial inverters. 

\textbf{Test-bed devices.}
To support the victim inverter's operation, we use a programmable solar panel emulator TEWERD TPV1000~\cite{TPV} to emulate solar panels and a RIGOL RP1025D high voltage differential probe~\cite{rp} to acquire the real voltage. 

\textbf{Adversary devices.} 
The adversary devices are the same as those introduced in Section~\ref{sc4}. They are used to generate, amplify, and emit EMI signals. \textit{To prevent the adversary devices from causing conducted interference to the victim's PV inverter through the public grid, we added a fourth-order low-pass filter between the adversary devices and the grid to eliminate conducted interference.}

\subsubsection{Evaluation of \dos}
\begin{figure}[t]
%	\vspace{-1em}
	\centering
	\subfigure[Result of \dos. \ding{172}: Before EMI, \ding{173}: EMI begins, \ding{174}: After EMI.]{\includegraphics[width=4.1cm]{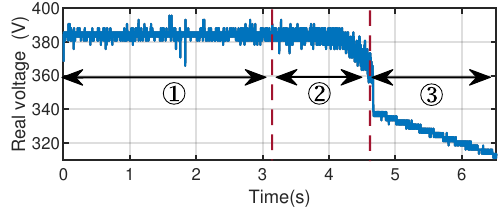}\label{sleep}}
        \hspace{2mm}
	\subfigure[Result of \damage. \ding{172}: Before EMI, \ding{173}: EMI begins, \ding{174}: Burning out, \ding {175}: After EMI.]{\includegraphics[width=4.1cm]{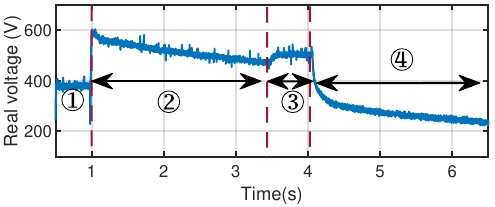}\label{die}}
	\caption{The experiment results of \dos and \damage.}
%	\vspace{-1em}
\end{figure}

We have introduced in~\Cref{sc5} that \dos impact can be induced in two ways: 

\textbf{\dos on the DC side.} Taking the TI C2000 inverter as an instance, we use a signal generator and RF amplifier to generate a signal with the frequency of $735  \, \mathrm{MHz}$ and the power of $10  \, \mathrm{W}$, and emit it with the antenna. As the measured $V_{dc}$ has been tampered with, we use the high-voltage probe to acquire the real $V_{dc}$, as shown in Fig.~\ref{sleep}. 

As we can see, before \dos, the PV inverter works correctly, and $V_{dc}$ remains stable at around $385  \, \mathrm{V}$. When EMI is initiated, we gradually increase the measured $V_{dc}$ to ``deceive'' the controller. As we presupposed, the controller reduces the real $V_{dc}$, and finally, the inverter shuts down at $4.5  \, \mathrm{s}$ due to current back-flow caused by under-voltage. The process can be seen in the video \textsuperscript{\ref{video}}.

\textbf{\dos on the AC side.} We first select the frequencies $1000  \, \mathrm{MHz}$ and $1080  \, \mathrm{MHz}$ that can respectively increase and decrease the measured AC voltage $V_{abc}$ through a frequency sweep. Then we generate EMI signal $s(t)$ by AM as described in Section~\ref{sc5}. The frequency of $s_m(t)$ is set to be the grid frequency of $50  \, \mathrm{Hz}$, and the total power is set to $10  \, \mathrm{W}$, although the selection of $s_m(t)$ is not unique. We can see that the ``Over-Grid Voltage'' alarm is triggered when the measured $V_{abc}$ increases to $240  \, \mathrm{V}$, and the ``Under-Grid Voltage'' alarm is triggered when the measured $V_{abc}$ is lower than $200  \, \mathrm{V}$ \textsuperscript{\ref{video}}.

The evaluation of \dos on commercial inverters are similar, and the result is shown in \Cref{result}. %Differently, 
As we can see, \ding{172} the \dos on commercial inverters need more EMI power, which is consistent with our observation: commercial inverters have better EMC countermeasures; \ding{173} the impact of \dos on the AC side is more challenging to achieve, compared with \dos on the DC side.

\subsubsection{Evaluation of \damage}
\damage can cause physical damages to the PV inverter by increasing the real $V_{dc}$. Through pre-test, we find that the $1350  \, \mathrm{MHz}$ EMI signal can reduce the measured $V_{dc}$. We adjust the total power from $5  \, \mathrm{W}$ to $20  \, \mathrm{W}$ and emit it with an antenna. We use the high-voltage probe to measure the real $V_{dc}$. 

The result is depicted in Fig.~\ref{die}. In phase \ding{172}, the PV inverter works correctly, and $V_{dc}$ remains stable at the target value of around $385 \, \mathrm{V}$. In phase \ding{173}, we emit EMI signal $s(t)$ and the controller increases the real $V_{dc}$ beyond $500  \, \mathrm{V}$. At around $3.5  \, \mathrm{s}$, the DC capacitor gets a dielectric breakdown and burns out after a few seconds. However, the PV inverter is ``unconscious'', and $V_{dc}$ continues to rise from $3.5  \, \mathrm{s}$ to $4  \, \mathrm{s}$. To prevent any danger, we terminate the test and cut off the power supply at $4  \, \mathrm{s}$ and the voltage $V_{dc}$ decreases to 0, as shown in video~\textsuperscript{\ref{video}}. 

\subsubsection{Evaluation of \damp}
Based on the analysis in~\Cref{sc5}, if the adversary is assumed to have feedback information such as the input voltage $V_{pv}$ and current $I_{pv}$, 
he can pose a greater threat by decreasing the maximum power quantitatively.
Here we focus on the scenario where no feedback information is available and evaluate the \damp impact based on the ``interference'' method.

For the C2000 PV inverter, we set the input power of the inverter to $80  \, \mathrm{W}$, and then transmit a $1350  \, \mathrm{MHz}$ EMI signal and alternately switch it on and off. We find that the inverter's power is reduced to $30  \, \mathrm{W}$ and cannot be automatically adjusted to $80  \, \mathrm{W}$ during the \damp ~\textsuperscript{\ref{video}}. This indicates that \damp can interfere with the MPPT algorithm and reduce the inverter's power by $62.5\%$. 

For commercial inverters, we set the same V-I curve with a maximum power point of $2000  \, \mathrm{W}$ in the PV emulator. In the usual case, they can work stably at $1980  \, \mathrm{W}$, $1995  \, \mathrm{W}$ and $1960  \, \mathrm{W}$. Then, we conduct the \damp with a total power of $20  \, \mathrm{W}$ and record the power according to the PV emulator. As shown in \Cref{result}, the power of Ginlong, Kstar, and Huawei PV inverters can be reduced by $150  \, \mathrm{W}$, $105  \, \mathrm{W}$ and $110  \, \mathrm{W}$ at most, respectively. Besides, we implemented the same experiment on GoodWe inverter~\cite{Goodwe50} under a real-world microgrid, and its power is reduced from $35.6  \, \mathrm{kW}$ to $2  \, \mathrm{kW}$. The difference in reducible power is mainly caused by the perturbation resistance of different MPPT algorithms and the difference between the PV emulator in the laboratory and the real PV panel in the real-world microgrid.

Compared with \dos, \damp can be more insidious in some sense. On the one hand, it can be utilized to affect the 
power conversion efficiency of PV generation in the long term; on the other hand, it can launch in an on/off pattern (i.e., switching attacks) to affect the PV microgrid, as discussed in \Cref{Exploitability}.
%BTW, as remarked and clarified, if the adversary can obtain feedback information, %the PV power can be controlled quantitatively.

%%%%%%%%%%%%%%%%%%%%%%%%%%%%%%%%%%%%%现在的表格%%%%%%%%%%%%%%%%%%%%%%%%%%%%%%%%%%%%%%%%%%%%%%%%%%%%%%%%%%%%
\begin{table*}[h]
	\vspace{-0.5em}
	\caption{Result of \alias on PV inverters.} 
	\begin{center}
		\setlength\tabcolsep{6pt} %调整列间距，即表格宽度
		\begin{threeparttable}
                \renewcommand{\arraystretch}{1.0} %调整行宽，调为原来的1.5倍
			\begin{tabular}{c|c|c|c|c|c|c|c|c|c|c|c|c|c}
				\toprule[2pt]
				\multirow{3}*{\makecell{Inverter }} & \multicolumn{6}{c|}{\dos} & \multicolumn{3}{c|}{\damage} & \multicolumn{4}{c}{\damp}  \\
				\cline{2-14}
				& \multicolumn{3}{c|}{On DC side} & \multicolumn{3}{c|}{On AC side} &\multirow{2}*{\makecell{Pow. \\ (W)}} &\multirow{2}*{\makecell{Freq.\\(MHz)}} &\multirow{2}*{Result}&\multirow{2}*{\makecell{Freq.\\(MHz)}} &\multirow{2}*{\makecell{Pow.(W)\\before \\ \damp}} & \multirow{2}*{\makecell{Pow.(W)\\after \\ \damp\tnote{2}}}& \multirow{2}*{\makecell{Pow.\\dev. rate}} \\
				\cline{2-7}
				&\makecell{Pow. \\ (W)}&\makecell{Freq.\\(MHz)}&\makecell{Success \\ rate\tnote{1}}&\makecell{Pow. \\ (W)}&\makecell{Freq.(MHz)\\Pos./Neg.\tnote{4}}&\makecell{Success \\ rate\tnote{1}} & & & & & & & \\
				\hline
				Ti C2000  & 5 & 735 & 100\%& 5 & 1036/1490 & 100\% & 10& 1000 & 100\% & 760 & 80 & 30 & 62.5\%\\
				\hline
				Ginlong  & 10 & 916 & 100\%& 10 & 625/1210 & 80\% & -\tnote{3}& - &- & 1192 & 1980 & 1830  & 7.6\% \\
				\hline
				Kstar  & 10 & 749 & 100\%& 10 & 990/810 & 90\% & -& - & - & 998& 1995 & 1890  & 5.3\% \\
				\hline
				Huawei\tnote{5}  & 10 & 1150 & 100\%& 10 & 980/1020 & 80\% & -& - &- & 1330 & 1960 & 1850  & 5.6\% \\
				\hline
				GW(LCD,50kW)  & 20 & 920 & 100\%& - & - & - & -& - &- & 960 & 35.6k & 2k  & 94.3\% \\
				\hline
				GW(LED,60kW)\tnote{6}  & 20 & 945 & 100\%& - & - & - & -& - &- & - & - & -  & - \\
				\bottomrule[2pt]
			\end{tabular}
		\begin{tablenotes}
			\footnotesize
			\item[1] For \dos on each inverter, we repeat the experiment 10 times from different angles and calculate the success rate.
			\item[2] For \damp, we repeat the experiment 10 times and calculate the average deviation.
			\item[3] ``-'' means there is no result~(Considering safety or other factors).
			\item[4] ``Pos./Neg.'' means the positive and negative deviation; ``Pow.'' means power; ``Freq.'' means frequency.
			\item[5] For Huawei SUN2000, all the experiments were done with the shell removed, we will give detailed reasons in \Cref{discussion}.
			\item[6] For Goodwe(LED,60kW) inverter, we don't perform \damp because there is no LCD screen to display the power.
		\end{tablenotes}
		\end{threeparttable}
		\label{result}
	\end{center}
	%\vspace{-0.5em}
\end{table*}

\subsection{Evaluation on PV Microgrid}
To demonstrate the threat of EMI to the real-world grid, we collaborate with the local distribution grid operator and conduct the \dos and \damp experiments on a real-world microgrid, ensuring safety and minimal disruption to residents' daily lives.%not affect residents' daily lives.

%微网的拓扑结构如图X所示，
The topology of the microgrid is shown in Fig.~\ref{jiegou} in~\Cref{supplementary}. The microgrid has a capacity of $400  \, \mathrm{kVA}$, and the maximum generated power of PV is $323  \, \mathrm{kW}$. The total load is usually between $12  \, \mathrm{kW}$ and $40  \, \mathrm{kW}$. To ensure a continuous and stable power supply, the microgrid is designed with a $150  \, \mathrm{kWh}$ battery energy storage (BES) system. It can operate in grid-connected or islanding mode, serving a discrete footprint of a remote mountain village. The PV microgrid contains 2 types of 5 PV inverters designed by GooDWe with the power of $50  \, \mathrm{kW}$ and $60  \, \mathrm{kW}$. 

\begin{figure}[t]
	%\vspace{-1em}
	\centering
	\subfigure[Experiment setup in the real-world microgrid.]{\includegraphics[width=3.4cm]{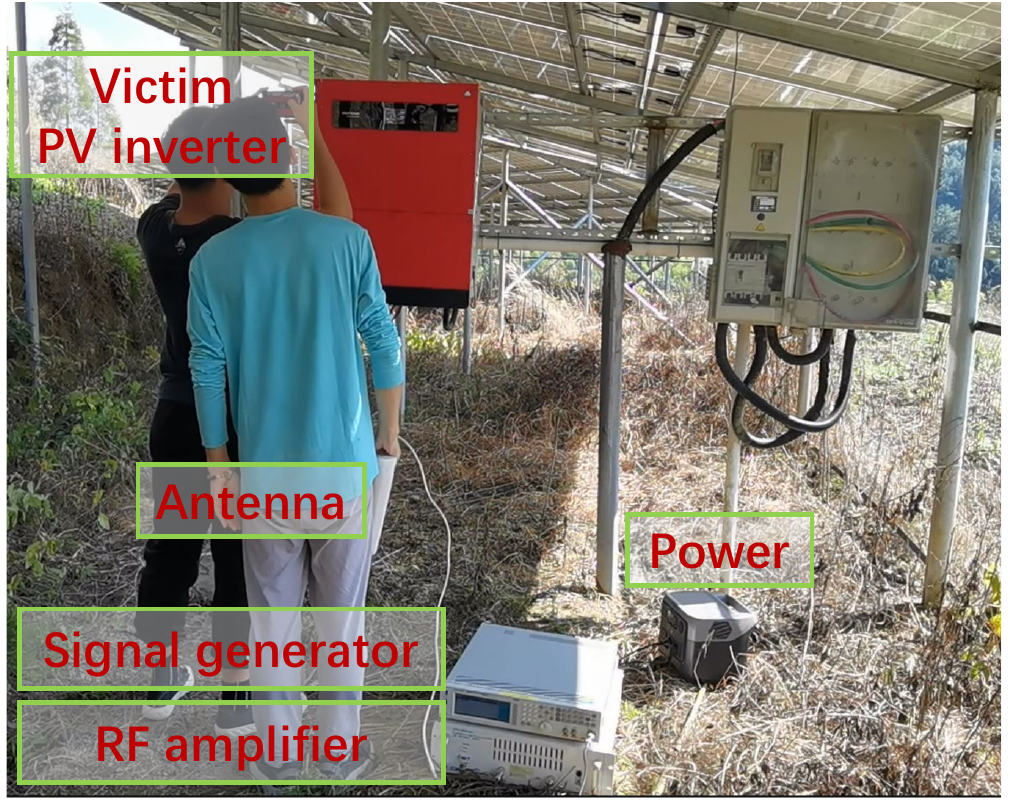}}
        \hspace{2mm}
	\subfigure[Impact of \dos on microgrid frequency.]{\includegraphics[width=5.0cm]{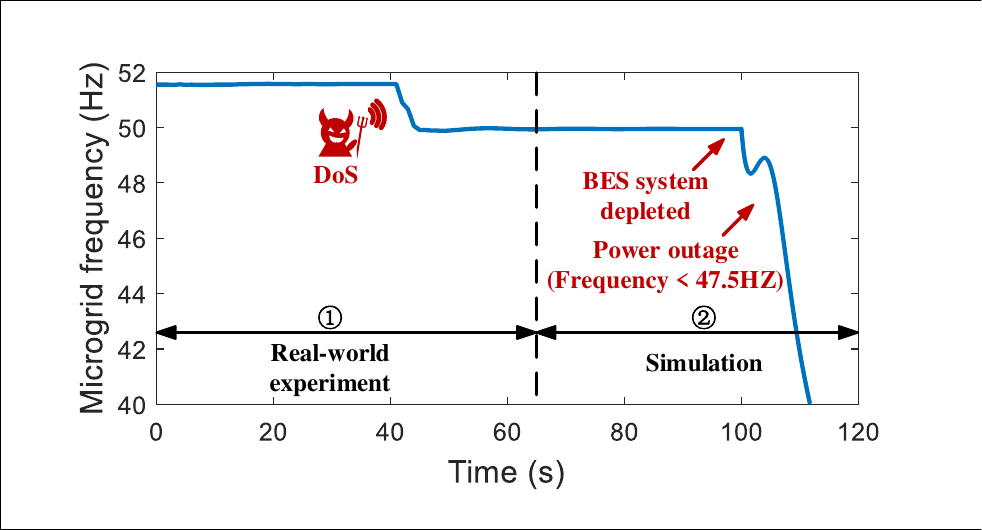}}
	\caption{The impact of \dos on a real-world PV microgrid's frequency. stage\ding{172} real-world experiment, stage\ding{173} simulation.}
    \label{microgridTest}
	%\vspace{-1em}
\end{figure}

In the islanding mode of the microgrid, we first evaluated \dos and \damp on each inverter. Then we perform \dos on all 5 PV inverters and lasts for around $1  \, \mathrm{min}$. We investigated the impact of the \dos on the islanding mode microgrid and recorded the frequency of the microgrid in~Fig.~\ref{microgridTest}. 

It can be observed that there is a decrease in the microgrid frequency by $1.5 \, \mathrm{Hz}$. This shift is caused by the deficiency of PV generation at the point, prompting the BES system from the P/Q control \cite{dong2018residential} to V/f control \cite{chen2011modeling}. The P/Q mode controls the output power of the PV-BES system, while the V/f mode controls the output voltage/frequency by the BES output. This indicates that the microgrid is now solely powered by the BES system, and the battery energy is continuously depleting. Notably, such a condition, mainly when the battery is low on energy, may cause more severe consequences. 

However, we are not permitted to conduct the experiments under %extreme 
conditions of %shallow
extreme low power storage that leads to over-discharging, as it could harm the health of the BES. Thus, we %entirely 
modeled the entire microgrid and simulated the consequence of \dos under insufficient energy storage in 
the simulator PowerWorld. As shown in~Fig.~\ref{microgridTest}, the battery in the BES system is depleted in the absence of PV input for a while, and the frequency of the microgrid decreases rapidly, leading to a power outage (according to the IEEE Std 1547-2003 \cite{ieee20111547}, in microgrids, the frequency deviation should not be greater than 5\% of nominal). Note that as long as the PV output power is less than the load power, the BES system will continue to discharge, ultimately leading to a power outage of the microgrid. %This also means that all three attacks we propose are capable of causing microgrid instability.

\subsection{Influence Quantification}
Based on the principle of EMI, the EMI distance and power can influence the \alias threat. In this subsection, we analyze the influence of EMI distance and power on \alias under the threat model. 
\subsubsection{Influence of EMI Distance and Power on Inverter Sensors}
Here, we evaluate the effects of \alias on the deviation of the DC bus voltage $\Delta V_{dc}$ at $0\sim 215  \, \mathrm{cm}$, using $5  \, \mathrm{W}$, $10  \, \mathrm{W}$, $20  \, \mathrm{W}$ and $50  \, \mathrm{W}$ as the total power. The result is depicted in Fig.~\ref{dp}(a). We can see that higher power allows for a greater working distance. Taking the C2000 PV inverter as an instance, the self-protection mechanism will be triggered when the $V_{dc}$ suddenly changes by $30  \, \mathrm{V}$. %In theory, 
With a $20  \, \mathrm{W}$ EMI device, the inverter can be affected at a distance of around $150  \, \mathrm{cm}$.

%\subsubsection{Influence of EMI Power on Inverter's Sensors}
We placed the antenna at distances of $50  \, \mathrm{cm}$ and $100  \, \mathrm{cm}$ from the target PV inverter and tested the effects of power on the deviation of the DC bus voltage $\Delta V_{dc}$. The result is shown in Fig.~\ref{dp}(b). %We assume 
For the adversary's target %is 
to generate a $30  \, \mathrm{V}$ offset on $\Delta V_{dc}$, when the distance is $50  \, \mathrm{cm}$, the adversary only needs an EMI power of $5  \, \mathrm{W}$.

\subsubsection{Influence of EMI Distance and Power to \dos the Commercial Inverter}
\begin{figure}[tb]
%	\vspace{-1em}
	\centering
	\subfigure[Distance$\to$sensor.]{\includegraphics[width=2.7cm]{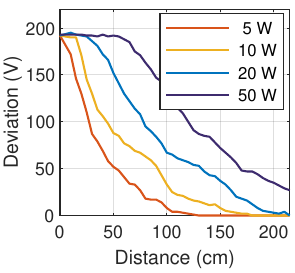}}
	\subfigure[Power$\to$sensor.]{\includegraphics[width=2.77cm]{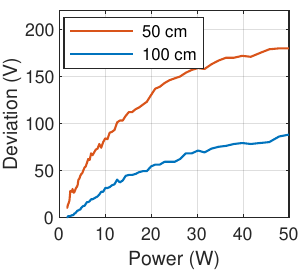}}
	\subfigure[Dis.\& Pow.$\to$ \dos.]{\includegraphics[width=2.77cm]{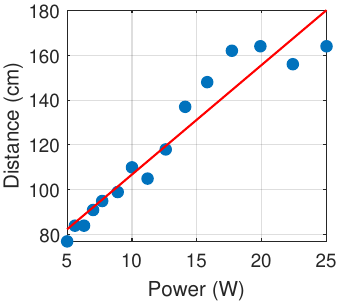}} \label{off}
	\caption{The influence of distance and power to manipulate inverter sensors and \dos a commercial inverter. The nonmonotonicity in (c) is mainly because the power will affect the electromagnetic field distribution of the antenna, which is not linear.}
	\label{dp}
	%\vspace{-0.5em}
\end{figure}

Since commercial inverters respond similarly to EMI,  we chose a well-selling commercial inverter, Kstar BluE-G, and recorded the maximum distance to perform \dos at a specific power. As shown in Fig.~\ref{dp}(c), we can see that a $20  \, \mathrm{W}$ EMI can achieve \dos at a distance of $160  \, \mathrm{cm}$, consistent with our threat model.

% !TEX root = ../Theremin.tex
% \clearpage
\section{Discussion}
\label{discussion}
In this section, we analyze the limits, diversity, 
and countermeasures of \alias, and design a portable and cost-effective EMI device to show its practicality. 

\subsection{Limits of \alias}
\subsubsection{Subject to Power and Distance}
The EMI power and distance are crucial impact factors of \alias. Essentially, \textit{our work represents one type of attacks exploiting analog signals. Such analog attacks have to follow the law of physics and a larger impact distance requires a more powerful transmitter.} Notably, we find that \dos has great upward compatibility with power. For example, if a $10 \, \mathrm{W}$ EMI signal
at $50 \, \mathrm{cm}$ can shut down the inverter, then EMI signals with $20 \, \mathrm{W}$, $30 \, \mathrm{W}$ or even $50 \, \mathrm{W}$ can achieve the same effect. %impact. 
The adversary shall choose the highest possible power for success. For exploitability, attackers can disguise themselves as a passerby or remotely control drones carrying our designed portable devices, as demonstrated in video \textsuperscript{\ref{video}}.
%damping对商用逆变器的效果有限

\subsubsection{Limited Impact Scale}
Different from cyber-attacks that may cause large-scale outages, the impact of our attack is limited to PV inverters and potentially local PV microgrids. For a larger-scale grid, there may be greater resilience to compensate for the PV power. Thus, for attackers with different goals, EMI may not always be the best approach. Besides, attackers with physical access to the inverter may launch simpler attacks with more predictable consequences. Nonetheless, EMI attacks can be stealthier than cyberattacks in terms of digital traces, and they are also safer for attackers compared with direct physical attacks. We believe \alias is applicable to local microgrid-scale attack scenarios where the attack needs to be stealthy and difficult to trace back.

\subsection{Diversity of \alias}

%既然alias能够实现对单个传感器或多个传感器的同时操控，这可以被攻击者利用来扩展更多类型的攻击，比如操控逆变器的输出频率、输出的功率因数等.这些更复杂的攻击也为控制算法的鲁棒性带来了更大的挑战，这需要被重视
\subsubsection{Diversity of the Impact}
%In this paper, 
We propose \dos, \damage, and \damp to illustrate the threat of \alias. Since EMI can control multiple sensors simultaneously, adversaries can use it to explore more impacts, such as controlling the output frequency, the output power factor, and more. For example, EMI can also introduce harmonics~(using the method in Fig.~\ref{siheyi}) into the AC output of the inverter and damage electrical appliances or devices.

\subsubsection{Diversity of the Victim}
This study highlights the vulnerability of op-amp-based voltage and current sensors in PV inverters to EMI. While PV inverter is a typical example of power electronic devices, the scope of potential victims can extend. Similar sensor technologies and energy conversion processes are prevalent in various applications, including power grids, electric vehicles, and industrial machinery. %, and consumer electronics. 
Additionally, the control algorithms employed in different inverters partly exhibit similar characteristics. For instance, the battery storage inverter may adopt the TSPC system \cite{liu2022review}, 
implying the presence of a DC bus capacitor in such inverters
and the associated impact of \damage and \dos. Consequently, it is imperative that the security analysis by \alias should also be performed in these diverse domains.

%\subsection{Potential scenario of \alias}
\subsection{Exploitability of \alias}
\label{Exploitability}
\begin{figure}[t]
	\centerline{\includegraphics[width=8cm]{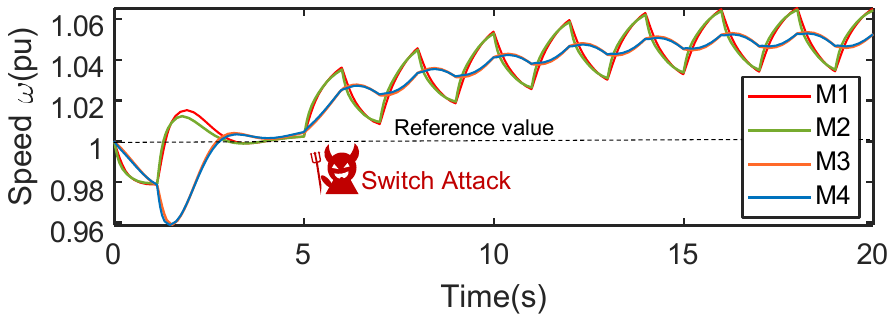}}\vspace{-2mm}
	\caption{The simulation result of Switching Attack with \damp.}
	\label{scenario}\vspace{-1mm}
\end{figure}

\alias may cause consequences to the microgrid that go beyond those achieved in our evaluation, under specific conditions where there are both solar PV and synchronous generators in a grid. Particularly, for the \damp that can manipulate the output power of PV inverter by more than 90\% (as tested in the real-world microgrid), it can launch in an on/off pattern and induce low-frequency oscillations of power supplies, which may cause physical damage of other synchronous generators and even result in a power outage, similar to how Switching Attacks~\cite{kirby2003frequency} affect the grids~\cite{ghafouri2022coordinated}. This is because, the low-frequency oscillations can result in angular speed oscillations of generators, which can lead to damage or disconnecting of the generators. It has been demonstrated that manipulating a mere 1.23\% of the total system power is enough to achieve the Switching Attack~\cite{hammad2017class}. To further verify, we simulate the use of \damp to oscillate the angular velocity of generators in the grid (the modified Kundur benchmark system with four synchronous generators and two PV farms \cite{Kundur}) via Simulink, and our simulation result shows that \damp could cause this cascading failure effectively, %in a minute, 
as shown in Fig.~\ref{scenario}.

\subsection{Countermeasures}
\subsubsection{Vulnerability of Filtering Leakage and Corresponding Countermeasures}

\textbf{\ding{172}~Filtering of Ultra-high Frequency~(UHF) Noise.} Low-pass filtering is often employed for the suppression of high-frequency interference. However, the low-pass filter in practice shows leakage at UHF noise, as shown in Fig.~\ref{shizhen}. 
%This is due to the loss of inductance and capacitance efficiency. 
For low-pass filters composed of inductors and capacitors, leakage occurs when the input signal frequency is much higher than the filter cut-off frequency~\cite{RFIFILTER}. 

A potential countermeasure is to design filters for different frequency bands of interference. The three-stage low-pass filter structure has better performance in dealing with high disturbance of UHF noise, with a low-frequency bandwidth of $10  \, \mathrm{kHz} \sim 1  \, \mathrm{MHz}$, medium frequency bandwidth of $1  \, \mathrm{MHz} \sim 100  \, \mathrm{MHz}$, and high-frequency bandwidth of $100  \, \mathrm{MHz} \sim 1  \, \mathrm{GHz}$ (Fig.~\ref{filter} in \Cref{supplementary}).

\textbf{\ding{173}~Shielding Against Ultra-high Frequency Noise.} The filtering effect of low-pass filter circuits on Ultra-high noise is limited by its physical characteristics. According to \Cref{result}, we have evaluated 5 off-the-shelf PV inverters, all of which have EMC-proof metal-case shielding. However, SUN2000 is the only inverter whose shielding is ``effective'' in resisting \alias within the threat model. To explore, we sawed and deconstructed 3 commercial PV inverters to compare their cases~(Fig.~\ref{thick} in \Cref{supplementary}) and found different thicknesses of metal cases ($4 \, \mathrm{mm}$ on SUN2000 and $2 \, \mathrm{mm}$ on others). All of them meet the IP65 standard requirement according to the International Organization for Standardization ISO 20653:2023~\cite{ISO} and International Electrotechnical Commission IEC 60529~\cite{IEC}: with a thickness of 2 mm and waterproof silicone gasketsand. To further investigate how to strengthen the enclosures, we simulated the shielding effectiveness of 4 kinds of common shielding metals in COMSOL. According to our simulation in Fig.~\ref{simulationsetup} and Fig.~\ref{simulationresult} of \Cref{simu}, all enclosures above $2\,\mathrm{mm}$ thickness can meet the requirements of aerospace and military, but thicker enclosures can achieve better effects, such as resisting intentional EMI attacks. Additionally, we find that the LCD screens and cable interfaces may serve as entry points for external EMI. In our evaluation, only Huawei inverter~\cite{HuaweiInverter} adopts an LED display and can resist EMI. Previous work~\cite{researchon} suggests that the number of holes should be increased and the size of each hole should be reduced. For the PV inverters, we recommend using LED displays instead of LCDs and designing the cable interface with a metal connector to better block EMI.

\subsubsection{Consistency Checking for Anomaly Detection}
Since the PV inverter is a relatively ``stable'' system, there is often a relation between the sensors inside the PV inverter, due to the properties of power electronics. %device). %, and between the sensors and the external weather, etc. 
It is difficult for an adversary to fully follow the hidden ``rules'' when tampering with multiple sensors. Therefore, manufacturers can design a %detection algorithm for 
consistency checking algorithm. %model 
%to determine whether there is manipulation on the sensors.

%The existing detection\cite{zhang2021smart} methods can be mainly divided into two types: model-driven and data-driven. The model-driven method refers to establishing its state space estimation model based on the system and then comparing the model output and the system output in real-time to determine whether the system is under attack. The limitation of this method is that it is very difficult to establish an accurate state space model for a complex system.
%Data-driven method is base on machine learning, so the construction of the data-driven approaches do not depend on the network topology. However, the data-driven approach also has the disadvantage that it requires a large amount of historical data as the raw material for machine learning.

%Of course, this anomaly detection-based countermeasures have two drawbacks, one is the need for real-time computing will consume a certain amount of computing power. Second, this detection is limited to scenarios where the attacker implements simple EMI attacks, and if the attacker can emulate the manipulation of multiple sensors in accordance with the relationship of each sensor during normal operation of the PV inverter, this detection method will be deceived
%\subsection{Control Strategy Vulnerability Repair}

The threat of \alias also exploits the vulnerability in the control algorithm, which can be avoided in the controller design. For instance, the strategy to ensure the same input and output power of the entire inverter is to keep the DC bus voltage unchanged. %In this way, 
The current %DC/AC level 
control algorithm has no power information input from the PV panel to the inverter, and hence the adversary can manipulate the voltage by spoofing the DC bus voltage sensor. Fig.~\ref{Voltage control loop after vulnerability repair} presents a solution. This method inputs the voltage/current of the DC side and the AC side into the voltage control loop to calculate the input and output power on the capacitor $C_{dc}$ and compares it with the capacitor voltage %value 
$V_{dc}$ received by the voltage sensor. If the changes are consistent, it %means 
indicates that the DC bus voltage sensor is not under threat; otherwise, %there exist anomalies and 
the %PV 
inverter should issue an alarm.

\subsection{Portable and Cost-effective EMI Devices}
\begin{figure}[t]
	%\vspace{-1em}
	\centering
	\subfigure[Portable device.
	]{\includegraphics[width=2.6cm]{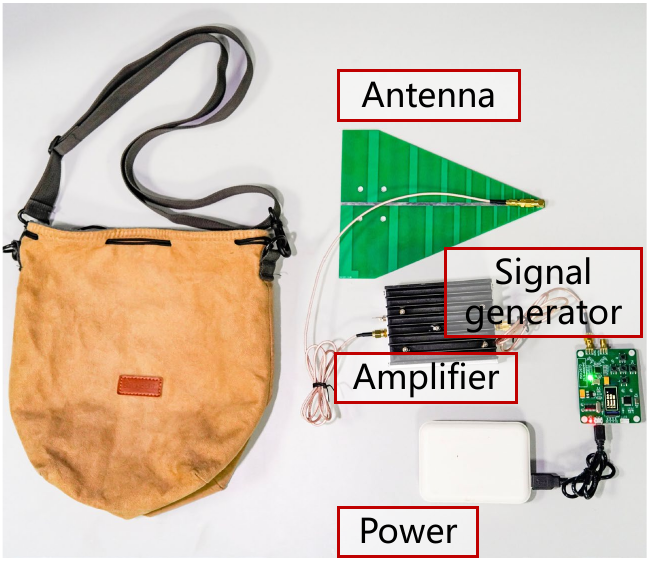}\label{mini}}
        \hspace{1mm}
	\subfigure[Vulnerability repair of voltage control.
	]{\includegraphics[width=5.8cm]{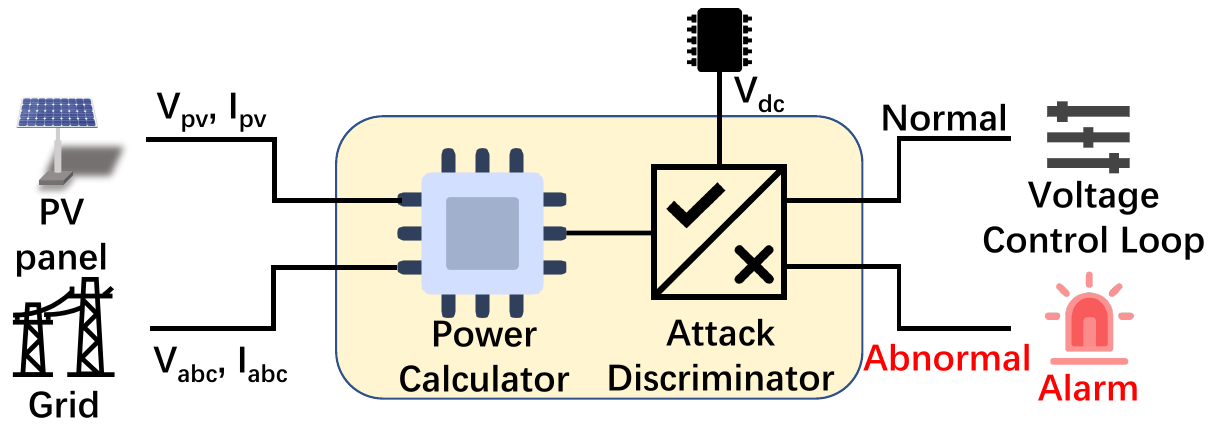}\label{Voltage control loop after vulnerability repair}}
	\caption{Portable  
		devices and control %logic 
		vulnerability repair.}
%	\vspace{-0.5em}
\end{figure}
We design a portable EMI device in Fig.~\ref{mini}. We select a mobile power %to support the entire system, 
with a maximum output of $5~V$ and $3~A$ and a cost of \$8. We select LTDZ MAX2870~\cite{ltdz} based on STM32 as the signal source that can generate signals in $23.5\sim 6000  \, \mathrm{MHz}$, %, which 
with a cost of %only 
\$45. Based on the result in Fig.~\ref{dp}(c), we choose a portable amplifier~\cite{Portable} with $10  \, \mathrm{W}$ power, costing \$95. %Finally, 
We then use an LS200-150 log-periodic antenna with a cost of \$9 to emit EMI signals. The entire device costs \$157 in total and can be hidden in a shoulder bag. Finally, we successfully use the portable device to implement \dos impact on commercial inverter~\cite{Kstar}; %, as shown 
see demo video \textsuperscript{\ref{video}}.

% !TEX root = ../Theremin.tex
% \clearpage
\section{Related works}
EMI attacks against sensors, actuators and communications have been studied extensively~\cite{ghost, kasmi2015iemi, xu2021inaudible, dai2023inducing, shoukry2013non, selvaraj2018electromagnetic, maruyama2019tap, wang2022ghosttouch, shan2022invisible, kohler2022signal, jiang2023glitchhiker}; the detailed comparison can be seen in \Cref{comparision} of \Cref{Comparision}. 
These best practices inspire us in the EMI attack design, such as determining the carrier frequency via frequency-sweep tests and modulating attack signals using amplitude modulation.
In comparison, 
\ding{172} to the best of our knowledge, we are the first to investigate the principle of achieving fine-grained incremental and decremental output control on differential op-amp-based sensors with EMI. We find and verify that the asymmetric layout of the differential input lines weakens the common-mode rejection function, and the amplified survival error can cause both incremental and decremental changes to the sensor output; 
\ding{173} we show the feasibility of manipulating complicated and robust control algorithms with EMI attacks. For instance, we propose new techniques for bypassing the Clarke/Park transformation and spoofing the MPPT algorithms. However, techniques commonly used in prior work such as injecting a constant deviation or random disruption to the sensor readings may fail to achieve \dos or \damp, as shown in Fig.~\ref{Girdsig_Vabc} of \Cref{cais} and Fig.~\ref{dmmo} of \Cref{grid};
\ding{174} we verify that the current inverters' enclosures need to be improved to cope with the intentional EMI concerns.

Current research on defense against EMI includes passive defense based on shielding~\cite{geetha2009emi, kondawar2020theory, thomassin2013polymer, wang2021polymer} and filtering~\cite{ye2004novel, wang2004effects, luo2012improving}, and active defense based on detection. Active defense includes 
\ding{172}~adding extra detection circuits to detect EMI~\cite{adami2014hpm, adami2011hpm, dawson2014cost, tu2021transduction, zhang2022detection}, 
\ding{173}~encoding critical signals secretly to detect EMI~\cite{kohler2022signal, ruotsalainen2021watermarking, shoukry2015pycra}, and 
\ding{174}~designing algorithm based on sensor characteristics to detect EMI~\cite{fang2022detection, kasmi2015iemi, kasmi2015automated, muniraj2019detection, ghosttouch}. The detailed comparison of these works is shown in \Cref{defense} of \Cref{Comparision}.
In conclusion, the passive defense methods can thoroughly eliminate the EMI threats with additional hardware costs; the active defense methods can typically detect \alias with lower costs, although there is a lack of effective ways to actively eliminate the impact of EMI after they are detected, which we believe is a direction of future work. 
%Millions of devices are expected to take proper EMI defenses in the future~\cite{zhang2023electromagnetic}. 

%Microphone: GhostTalk (S&P’13), Remote command Injection (IEEE EMC’15 and ANSSI’18), Inaudible Attack (IEEE TMTT’21),
%Magnetic Sensor: Non-invasive Spoofing Attacks For Anti-lock Braking Systems (CHES’13)
%Light Sensor: Electromagnetic Induction Attacks Against Embedded Systems (AsiaCCS’18)
%Temperature Sensor: Trick or heat? (CCS’19)
%Hall Sensor: Hall Spoofing (USENIX’20)
%Touchscreen: Tap’n Ghost (S&P’19), GhostTouch (USENIX’22), Wight (S&P’22), Invisible Finger (S&P’22) 
%CCD and CMOS Sensors: Signal Injection Attack against CCD Sensors (AsiaCCS’22)
%Smart Lock: Towards wireless spiking of smart locks (SPW’22)

%Many researchers have explored the possibility of using EMI to attack CPS. Denis et al.~\cite{ghost} evaluated the sensitivity of simulated sensor systems to signal injection attacks; G{\"o}k{\c c}en  et al.~\cite{pwm} demonstrated using EMI to attack digital signals; Selvaraj et al.~\cite{adc} use EMI signals attack anolog-digital and digital-anolog converters~(ADC and DAC) of embedded systems. However, one difficulty in EMI attack is how to make the injected weak signals perform a big role. To solve this problem, this paper systematically analyzes the vulnerability of GTIs' sensors and also control algorithms, and find a cost-effective attack signal transmission link that requires only 5 watts of power to ``shut down'' kilowatt GTIs.
% !TEX root = ../Theremin.tex
% \clearpage

\section{Conclusion}
In this paper, we systematically analyze the security of PV inverters and reveal the threat of EMI on both voltage and current sensors of PV inverters. We analyze the threat \alias and identify three destructive impacts, \dos, \damage, and \damp, which can cause the victim PV inverter to shut down, physically burn out, and reduce output power, respectively. The evaluations are successfully conducted on an inverter development kit, 5 off-the-shelf PV inverters and a real-world microgrid. Finally, we discuss the limits, diversity, exploitability and countermeasures of the threat. We hope our work can raise awareness of the security of power electronic devices in the grids with increasing RES. The observations could also facilitate the security analysis of other types of power electronic devices.


\section*{Acknowledgment}
We thank the anonymous reviewers for their valuable comments. This work is supported by the National Natural Science Foundation of China under Grant 62201501, 52161135201, 62201503, 61925109, 62222114, and 62071428.

%\newpage
\bibliographystyle{IEEEtranS}
\bibliography{sample_base}

% !TEX root = ../Theremin.tex
\appendix
%%%%%%%%%%%%%%%%%%%%%%%%%%%以上删掉%%%%%%%%%%%%%%%%%%%
\subsection{Control Algorithms in Simulations} 
\label{cais}
\begin{figure}[H]
	\centering
	\subfigure[Control algorithm of PLL.]{\includegraphics[width=3cm]{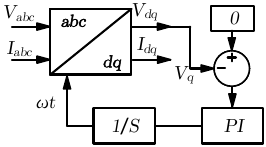}\label{PLL}}
	\quad
	\subfigure[Control algorithm of voltage control loop.]{\includegraphics[width=3cm]{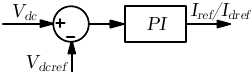}\label{VCL}}
	\caption{Control algorithms of PLL and voltage control.}
\end{figure}

\begin{figure}[H]
	\centering
	\subfigure[Control algorithm of single-phase current control loop.]{\includegraphics[width=2.8cm]{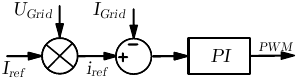}\label{sig_CCL}}
	\quad
	\subfigure[Control algorithm of three-phase current control loop.]{\includegraphics[width=4.8cm]{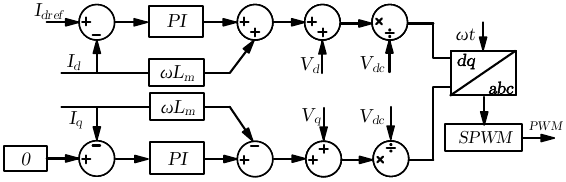}\label{three_CCL}}
	\caption{Control algorithms of the single-phase and three-phase current control loop.}
\end{figure}

This section gives the control algorithm in simulations, where Fig.~\ref{PLL} shows the control algorithm for PLL, Fig.~\ref{VCL} shows the voltage control loop, Fig.~\ref{sig_CCL} shows the current control loop for single-phase PV inverters, Fig.~\ref{three_CCL} shows the current control loop for three-phase PV inverters.

\begin{figure}[H]
	\begin{minipage}[tb]{0.45\linewidth}
		\centering
		\includegraphics[width=4cm]{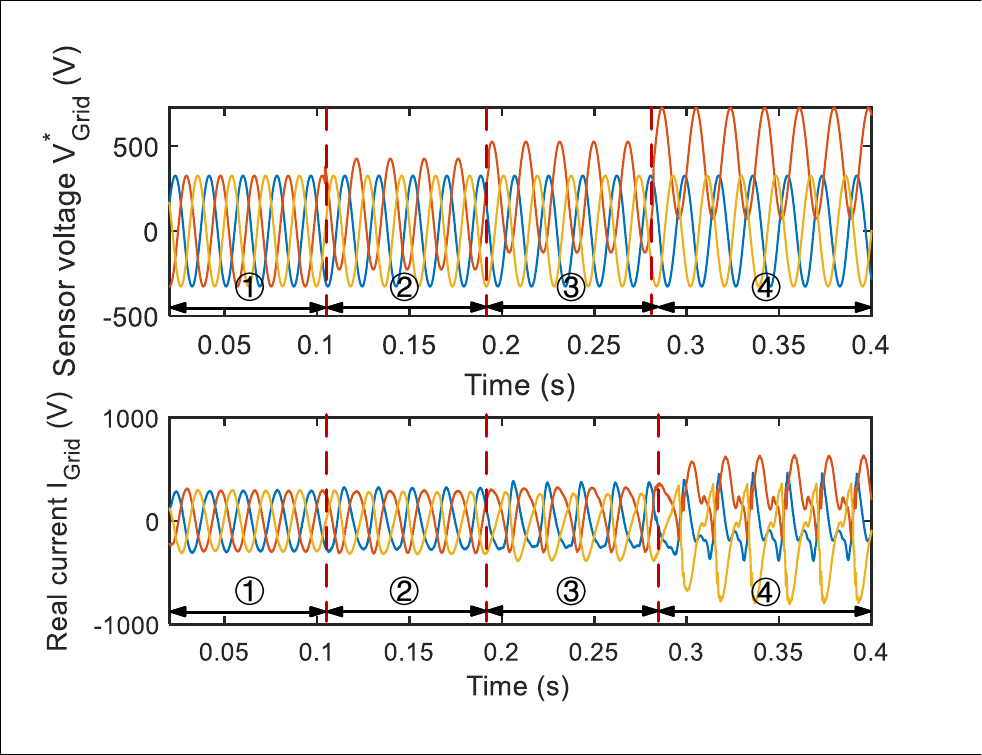}
		\caption{Inject the signal into one phase of the voltage sensor. \ding{172}:$V_a= 0V$, \ding{173}:$V_a= 50V$, \ding{174}:$V_a= 100V$, \ding{175}:$V_a= 200V$.}
		\label{Girdsig_Vabc}
	\end{minipage}
        \hspace{5mm}
        \begin{minipage}[tb]{0.45\linewidth}
		\centering
		\includegraphics[width=4cm]{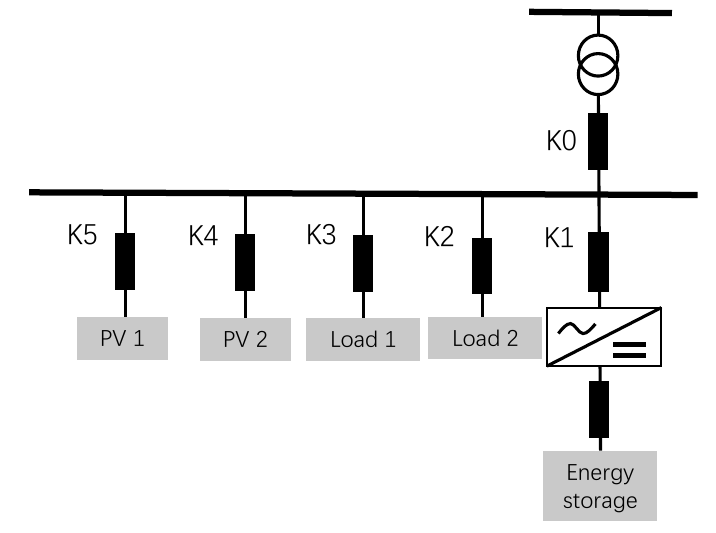}
		\caption{The structure of the micro-grid we evaluated.}
		\label{jiegou}
	\end{minipage}%
\end{figure}

\subsection{Simulation of Grid Sensor Tampering} \label{grid}
\subsubsection{Grid Voltage Sensor Tampering}
\label{Gridvoltagetampering}
We have given the simulation result of manipulating the current sensor in Fig.~\ref{gridtamper}. Here we introduce the simulation result of manipulating the grid voltage sensor in this section. 
\Cref{Girdsig_Vabc} gives the impact by injecting a constant signal into one phase of the voltage sensor, the effect is similar to that when manipulating the current sensor. Note that the effect is still reflected in the grid current since the grid voltage is unaffected by controls (grid voltage is generally stable and not controlled by individual inverters).

\begin{figure}[tb]
	\centering
	\subfigure[DC bus voltage. \ding{172}:$I_a= 0A$, \ding{173}:$I_a= 50A$, \ding{174}:$I_a= 100A$, \ding{175}:$I_a= 200A$, \ding{176}:$I_a= 400A$.]{\includegraphics[width=4.1cm]{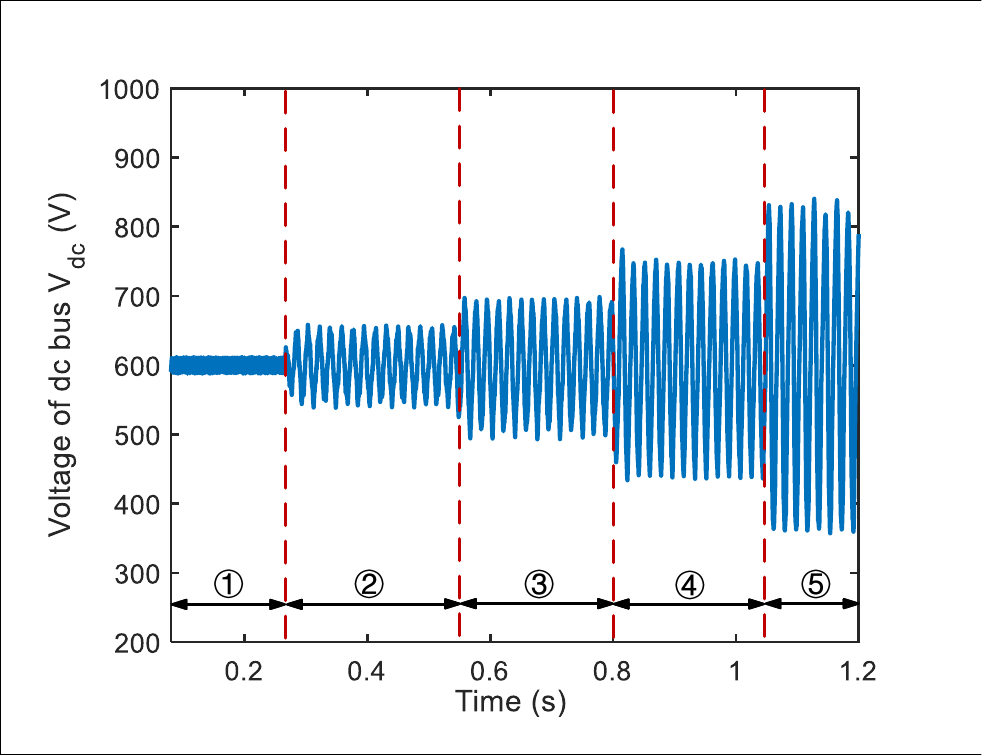}\label{Grid_DC}}
	\ \ 
	\subfigure[Active/reactive power output. \ding{172}:$I_a= 0A$, \ding{173}:$I_a= 50A$, \ding{174}:$I_a= 100A$, \ding{175}:$I_a= 200A$, \ding{176}:$I_a= 400A$.]{\includegraphics[width=4.1cm]{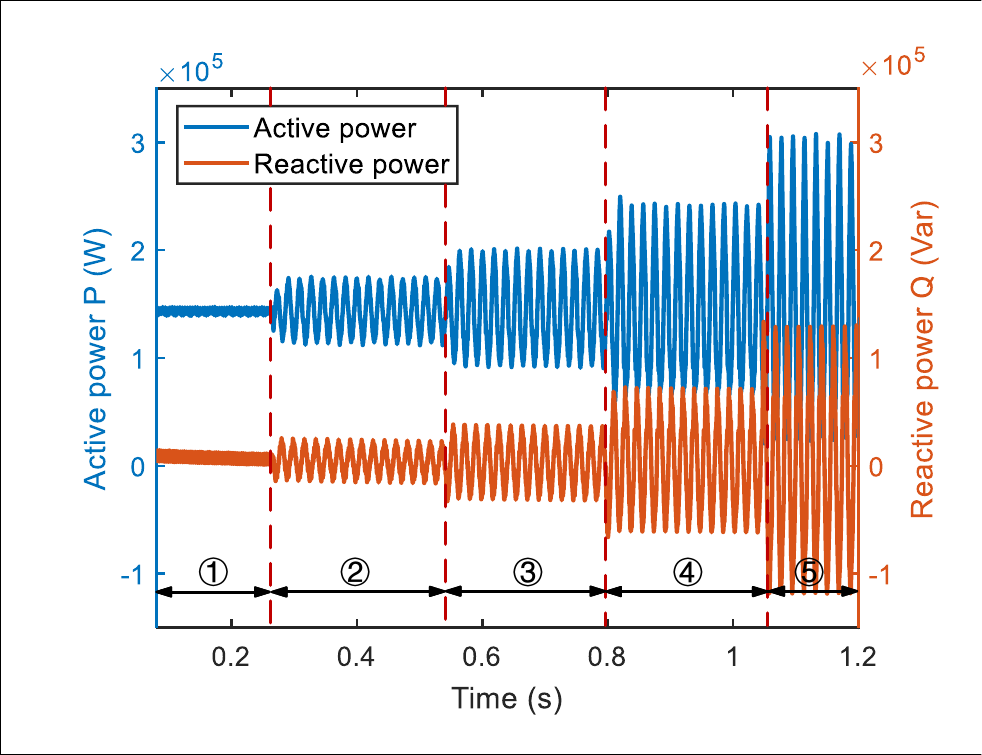}\label{PQ}}
	\caption{The effect of system oscillations. Effect of system oscillations on DC bus voltage and output active/reactive power when the inverter protection is not set or is not triggered.}
	\vspace{-1em}
\end{figure}

\begin{figure}[tb]
    \vspace{-0.5em}
	\centering
	\subfigure[Constant $\Delta I$ injection.]{\includegraphics[width=2.65cm]{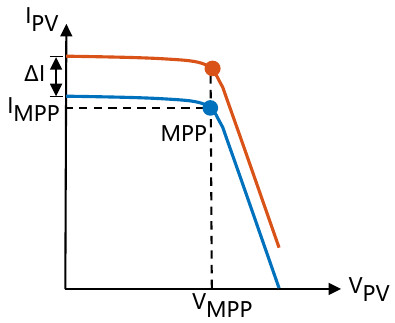}\label{PVinject1}}
	\subfigure[Constant $\Delta V$ injection.]{\includegraphics[width=2.65cm]{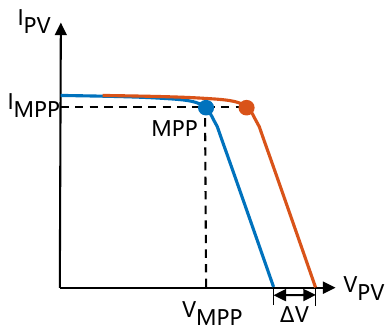}\label{PVinject2}}
	\subfigure[Fake V-I curve injection.]{\includegraphics[width=2.65cm]{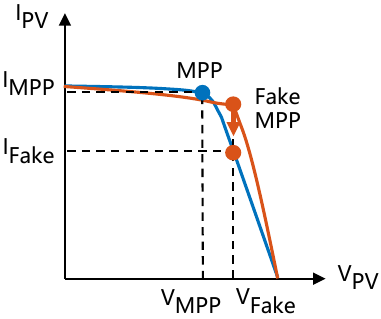}\label{PVinject3}}
	\caption{Different measurement manipulations on the V-I curve: (a) Inject a constant $\Delta I$ on the current measurement (b) Inject a constant $\Delta V$ on the voltage measurement;(c) Inject $\Delta I$ and $\Delta V$ by a specially designed V-I curve.}
        \label{dmmo}
   	\vspace{-1em}     
	\end{figure}

 \begin{figure}[tb]
	\begin{minipage}[tb]{0.45\linewidth}
		\centering
		\rotatebox{90}{\includegraphics[width=3cm]{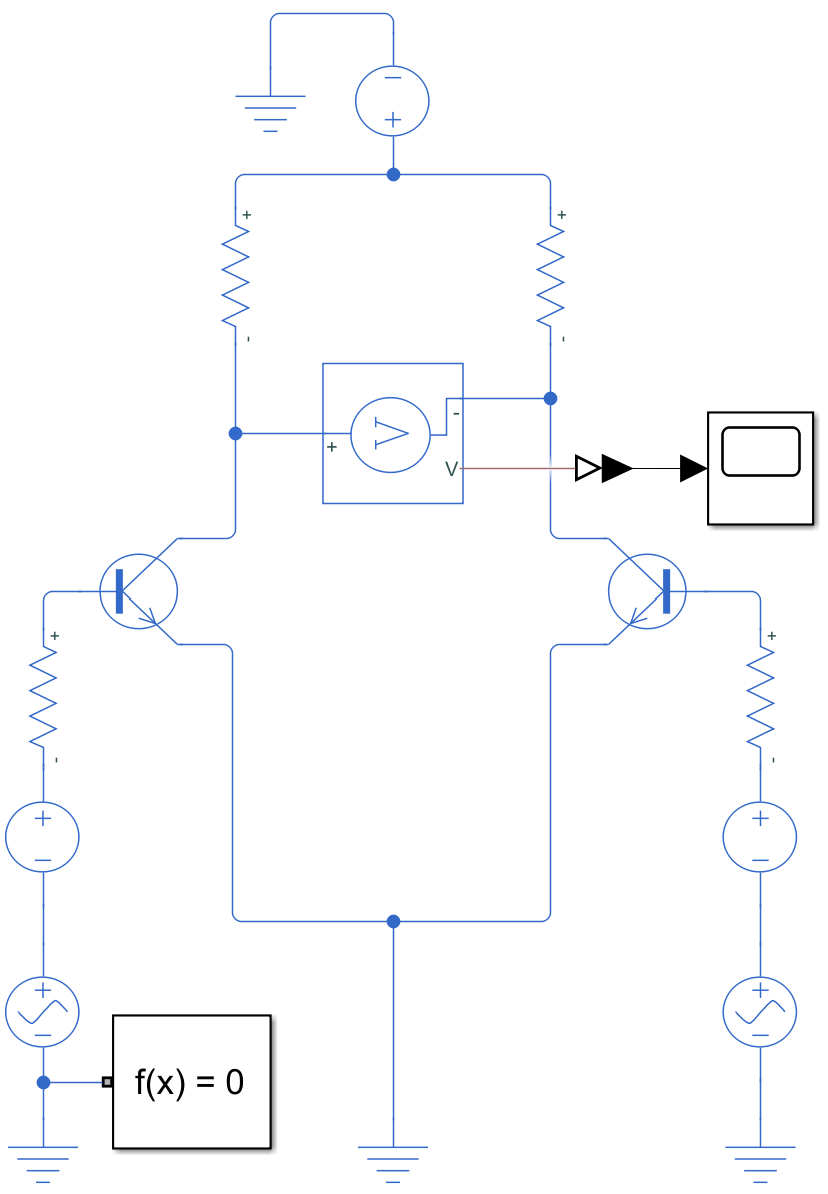}}
		\caption{The simulation model of the differential amplification input stage of op-amp chip in Simulink.}
		\label{model}
	\end{minipage}
        \hspace{5mm}
        \begin{minipage}[tb]{0.45\linewidth}
		\centering
		\includegraphics[width=4cm]{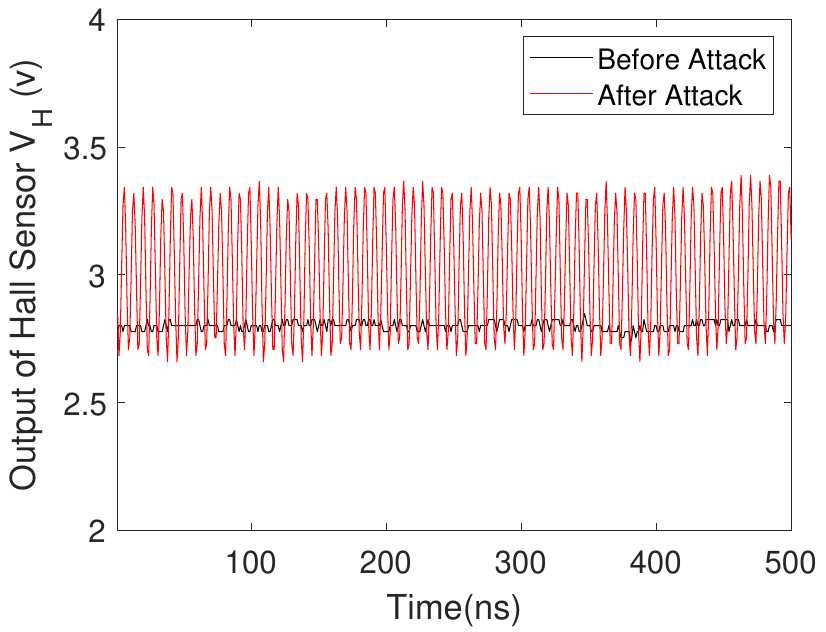}
		\caption{The output of Hall current chip under EMI.}
		\label{con}
	\end{minipage}%
\end{figure}

\subsubsection{Effect of Not Triggering Inverter Protection}
The impact of system oscillations on the inverter and the grid under EMI on the grid voltage/current sensor is given in this section, considering that some inverters do not have adequate protection mechanisms in place or that the protection is not triggered under EMI. The specific manifestations of system oscillations under EMI on the inverter are two main aspects:

\textbf{DC bus voltage.} The DC bus voltage $V_{Bus}$ oscillates violently, as shown in Fig.~\ref{Grid_DC}. Due to the system oscillation, the inverter output power is unstable, resulting in the DC bus capacitor $C_{Bus}$ constantly charging and discharging, which cannot be stabilized at the reference voltage.

\textbf{Output power.} The output active power oscillates and reactive power is injected into the grid, as shown in Fig.~\ref{PQ}. The quality of grid-connected power drops significantly.

\subsection{Supplementary Experimental Results}\label{supplementary}	
%\subsection{Simulation and Experiment of EMI on Voltage and Current Sensors}\label{hallcur}	
\textbf{Simulation and experiment of EMI on voltage and current sensors.} The simulation model of the differential amplification input stage of the op-amp chip is shown in Fig.~\ref{model}. The result of the EMI on Hall chip is shown in Fig.~\ref{con}.

%%%%%%%%%%%%%%%%%%%%%%%%%%%%%%%%%%%%%%%%%%%%%%%%%%%%%

\begin{table*}[tb]
	\caption{Comparison with previous EMI works}
	\begin{center}
            \setlength\tabcolsep{4.2pt} %调整列间距，即表格宽度
                \renewcommand{\arraystretch}{0.9}
			\begin{tabular}{c|c|c|c|c|c|c|c|c}
				\toprule[2pt]
	             \multirow{2}{*}{Work} & \multicolumn{2}{c|}{Attack Target} &  \multicolumn{3}{c|}{Attack Method} & \multicolumn{3}{c}{Attack Capability} \\ 
                    \cline{2-9}

                     &Victim & \makecell{Signal\\Type}  & \makecell{Attack Frequency} & \makecell{Utilize Victim's\\ Amplifier} & \makecell{Analyse Control \\ Algorithms} & \makecell{Attack \\ Power~$(\mathrm{W})$} & 
                     \makecell{Attack \\ Distance~$(\mathrm{m})$}& \makecell{Attack Effect} \\
                     \hline

                     \cite{ghost}& Microphone & Analog & $826\,\mathrm{MHz}$ & \ding{51} & \ding{55}& $10$ & $1 \sim2$ & Voice Command Injection\\  
                    \hline
                    
                    \cite{xu2021inaudible}& Microphone & Analog & $8  \sim16\, \mathrm{GHz}$ & \ding{51} & \ding{55} & $2.5$ & $2.5$& Voice Command Injection\\  
                    \hline
                    
                     \cite{shoukry2013non}& Magnetic sensor  & Digital & $500\,\mathrm{Hz}$ & \ding{55} & \ding{55}  & / & /&\makecell{Spoof wheel speed sensor}\\  
                    \hline
                    
                      \cite{selvaraj2018electromagnetic} & \makecell{Embedded system} & Digital & $170 \sim320\,\mathrm{MHz}$ & \ding{55} & \ding{55} &$1.8$ & $10$ &\makecell{Manipulated analog and digital signals}\\  
                    \hline
                     
                    \cite{trick} & Temperature sensor & Analog & $810 \sim950\,\mathrm{MHz}$ & \ding{51} & \ding{55}& $3$& $3$  &\makecell{Manipulate temperature sensors}\\  
                    \hline
                    
                    \cite{maruyama2019tap} & Touchscreen & Digital  & $60\sim90\,\mathrm{kHz}$ & \ding{55} & \ding{55} & $6$& $0.02$& Manipulate touchscreen\\  
                    \hline
                    
                    \cite{ghosttouch} & Touchscreen& Digital & $46\sim86\,\mathrm{MHz}$ & \ding{55} & \ding{55}& / & $0.04$  &Manipulate touchscreen \\  
                    \hline
                    
                    \cite{shan2022invisible} & Touchscreen& Digital & $140\,\mathrm{kHz}$ & \ding{55} & \ding{55}  & / & $0.7\sim2$& Manipulate touchscreen \\ 
                    \hline
                    
                    \cite{kohler2022signal} & CCD image sensor& Digital& $190\,\mathrm{MHz}$ & \ding{55} & \ding{55}  & $0.1$& $0.3$ &\makecell{Manipulate CCD image sensor}\\  
                    \hline

                    \cite{jiang2023glitchhiker} & \makecell{Camera signal line} & Digital & $1\,\mathrm{GHz}$ & \ding{55} & \ding{55} & / & $0.3$ &\makecell{Manipulate camera \\signal transmission} \\  
                    \hline

                    \cite{mohammed2022towards} & \makecell{Smart lock} & Digital & $500\,\mathrm{kHz}$ & \ding{55} & \ding{55} & / & $0.05$ &\makecell{Unlock the smart lock}\\  
                    \hline
                    
                    \cite{kohler2022brokenwire} & Charging system & Digital & / & \ding{55} & \ding{55} & $1$ & $47$ &\makecell{DoS communication \\ between charger and vehicle}\\  
                    \hline
                    
                     \cite{xie2023bitdance} & UART serial & Digital & $15.36\,\mathrm{MHz}$ & \ding{55} & \ding{55}& / & $0.05$ & UART signal bit flip\\
                    \hline
                    
                    \cite{pwm} & Servo & Digital & $8\sim140\,\mathrm{MHz}$ & \ding{55} & \ding{55} & $20$& $0.5$ & DoS \& Control servo\\  
                    \hline
                    
                     Our work & PV inverter& Analog& $735\sim1150\,\mathrm{MHz}$ & \ding{51} & \ding{51} & $20$ & $1.5$ & DoS \& Damage \& Damp\\   
				\bottomrule[2pt]
			\end{tabular}
		\label{comparision}	 
	\end{center}
\end{table*}

%%%%%%%%%%%%%%%%%%%%%%%%%%%%%%%%%%%%%%%%%%%%%%%%%%%%%%%%%%%%%%%%

\begin{table*}[t]
	\caption{Comparison of different defense methods against EMI attacks.}
	\begin{center}
            \setlength\tabcolsep{8pt} %调整列间距，即表格宽度
                \renewcommand{\arraystretch}{0.8}
			\begin{tabular}{c|c|c|c|c|c|c|c}
				\toprule[2pt]
                    \multicolumn{2}{c|}{\textbf{Type}} & \textbf{Work(s)} & \textbf{Detect EMI} & \textbf{Resist EMI} & \textbf{\makecell{No \\Additional \\Circuit}} & \textbf{\makecell{No \\Additional \\Computation}} & \textbf{\makecell{No \\Additional \\Cost}} \\ \hline
                    \multicolumn{1}{c|}{\multirow{3}{*}{\makecell{Active Defense \\(Detect EMI)}}} & \makecell{Add extra \\detection circuit} & \cite{adami2014hpm, adami2011hpm, dawson2014cost, tu2021transduction, zhang2022detection} & \ding{51} & \ding{55} & \ding{55} & \ding{51} & \ding{55} \\ \cline{2-8} 
                    \multicolumn{1}{c|}{}                   & \makecell{Encode critical \\signals secretly} & \cite{kohler2022signal, ruotsalainen2021watermarking, shoukry2015pycra} & \ding{51} & \ding{55} & \ding{51} & \ding{55} & \ding{51} \\ \cline{2-8} 
                    \multicolumn{1}{l|}{}                   & \makecell{Design algorithm\\ based on sensor \\characteristics} & \cite{fang2022detection, kasmi2015iemi, kasmi2015automated, muniraj2019detection, ghosttouch} & \ding{51} & \ding{55} & \ding{51} & \ding{55} & \ding{51} \\ \hline
                    \multicolumn{1}{c|}{\multirow{2}{*}{\makecell{Passive Defense \\(Shielding and Filtering EMI)}}} & \makecell{Shielding EMI} & \cite{geetha2009emi, kondawar2020theory, thomassin2013polymer, wang2021polymer} & \ding{55} & \ding{51} & \ding{51} & \ding{51} & \ding{55} \\ \cline{2-8} 
                    \multicolumn{1}{c|}{}                   & \makecell{Filtering EMI} & \cite{ye2004novel, wang2004effects, luo2012improving} & \ding{55} & \ding{51} & \ding{55} & \ding{51} & \ding{55} \\ 
				\bottomrule[2pt]
			\end{tabular}
		\label{defense}	 
	\end{center}
\end{table*}

%\subsection{The Victim PV Inverter, Sensors and Microgrid}

\begin{figure}[tb]
	\centerline{\includegraphics[width=8.9cm]{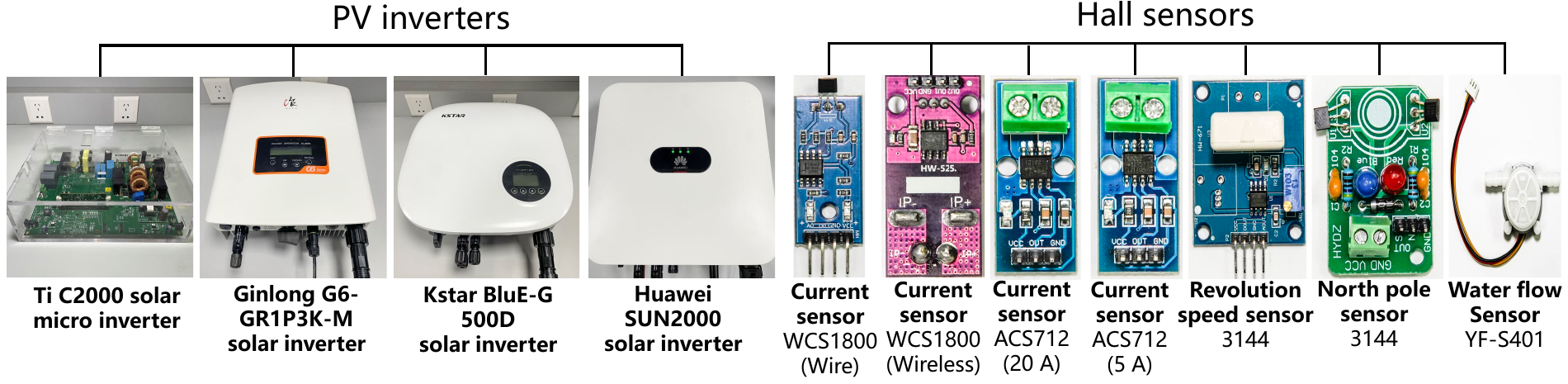}}\vspace{-3mm}
	\caption{The victim inverters and sensors we evaluated.}
	\label{zhaopian}%\vspace{-5mm}
\end{figure}

%\label{eva}
\textbf{The victim inverter, sensors and microgrid.} The tested off-the-shelf PV inverters and Hall sensors are shown in Fig.~\ref{zhaopian}. The structure of the tested micro-grid is shown in Fig.~\ref{jiegou}. 

%\begin{figure}[h]
%	\centering{\includegraphics[width=5cm]{figure/contrast.pdf}}
%	\caption{The EMI magnetic field superposition on Hall chip.}
%	\label{con}
%\end{figure}

%\subsection{The Filter Leakage and Corresponding Countermeasures}\label{leakage}

\begin{figure}[tb]%[b]
	\begin{minipage}[t]{0.45\linewidth}
		\centering
		\includegraphics[width=2.5cm]{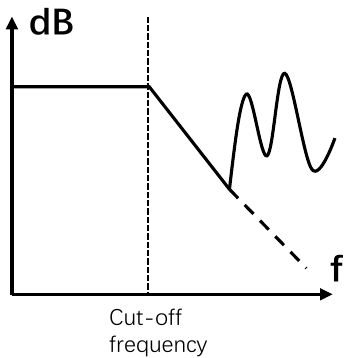}
		\caption{Leakage of the single-stage low-pass filter.}
		\label{shizhen}
	\end{minipage}%
        \hspace{5mm}
	\begin{minipage}[t]{0.45\linewidth}
		\centering
		\includegraphics[width=4cm]{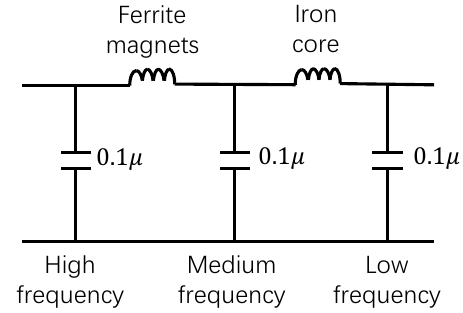}
		\caption{Multi-stage filter.}
		\label{filter}
	\end{minipage}
\end{figure}

\textbf{The filter leakage and corresponding countermeasures.} We show the filter leakage of a low-pass filter in Fig.\ref{shizhen} and a potential countermeasure using a multi-order filter in Fig.\ref{filter}.

\begin{figure}[H]
	\centerline{\includegraphics[width=6.2cm]{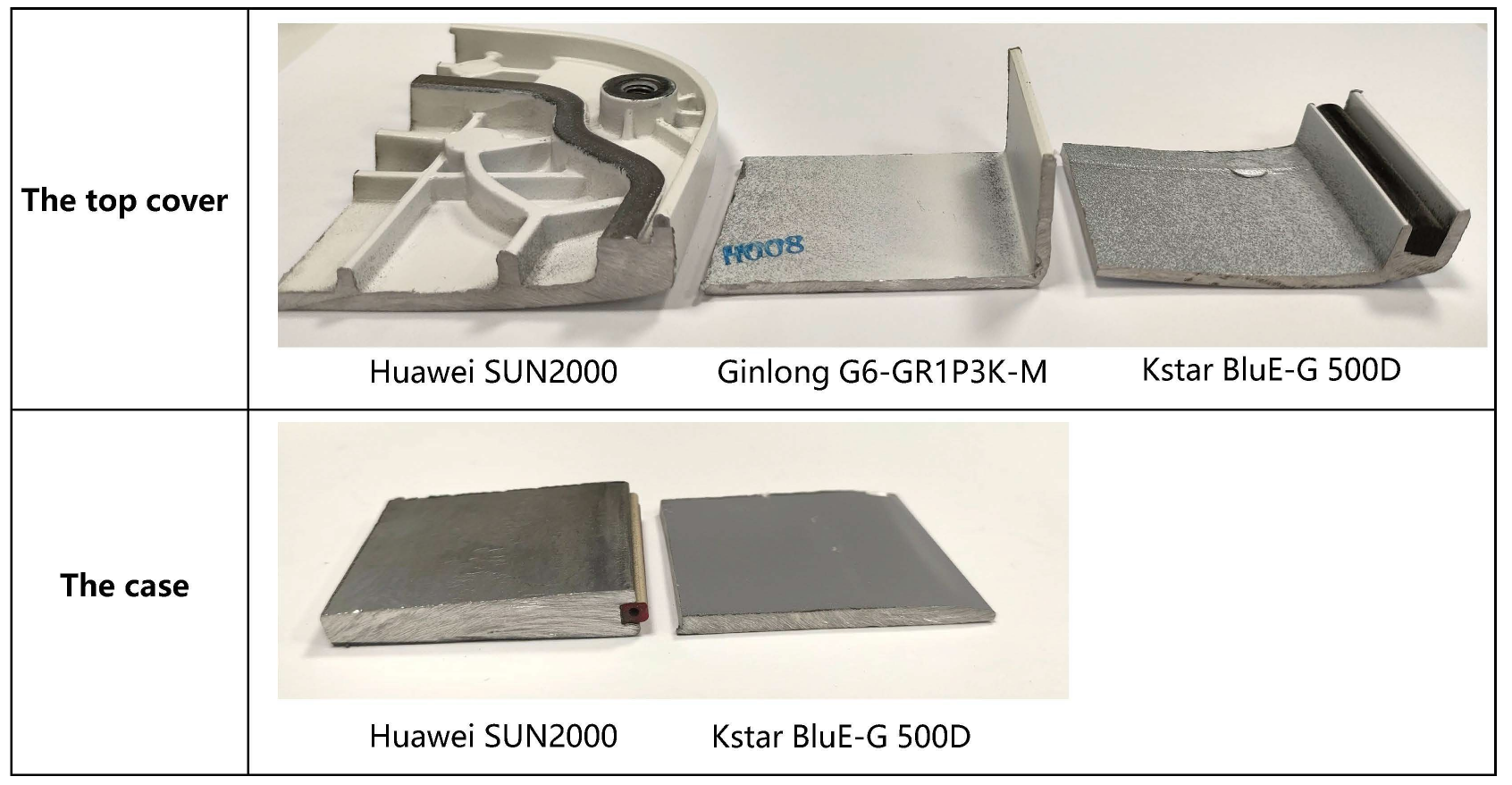}}%\vspace{-3mm}
	\caption{The thickness of 3 commercial inverters' cases.}
	\label{thick}%\vspace{-5mm}
\end{figure}

%\subsection{The Comparison of Cases of 3 Commercial PV Inverters}
\textbf{The comparison of cases of 3 commercial PV inverters.} We display the result of the disassembly and comparison of the cases of 3 commercial PV inverters in Fig.~\ref{thick}. 

\begin{figure}[tb]
	\centering
	\subfigure[The enclosure.]{\includegraphics[width=2.65cm]{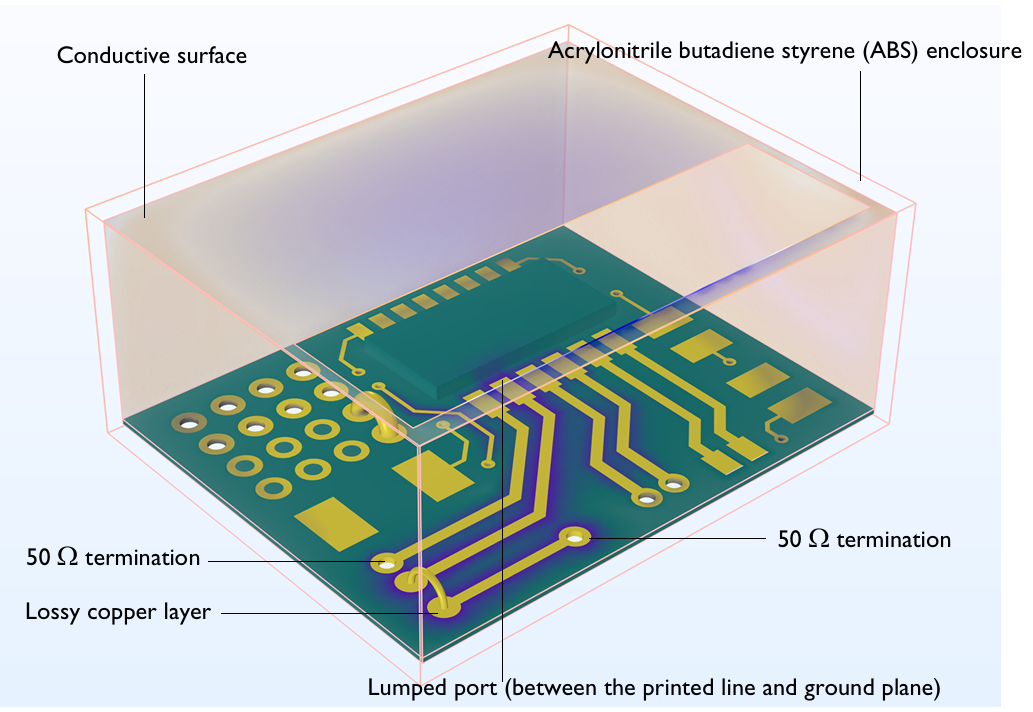}\label{pp1}}
	\subfigure[The placement.]{\includegraphics[width=2.65cm]{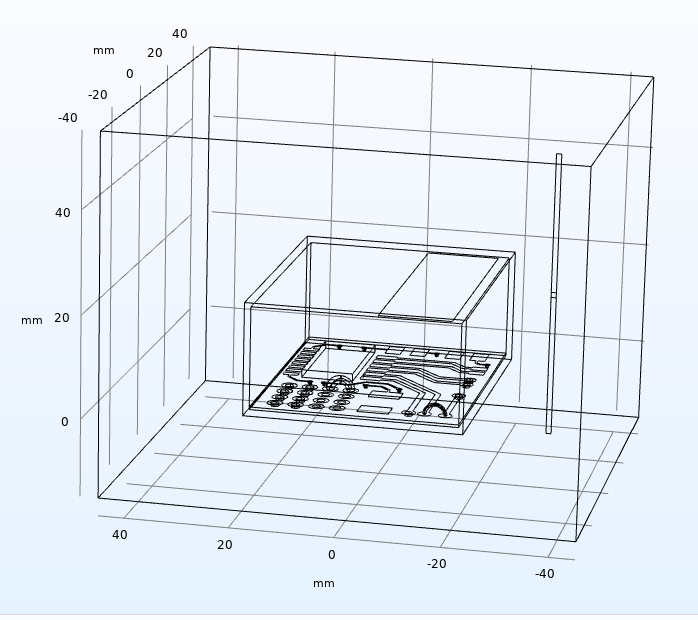}\label{pp2}}
	\subfigure[The EMI noise test points on PCB.]{\includegraphics[width=2.65cm]{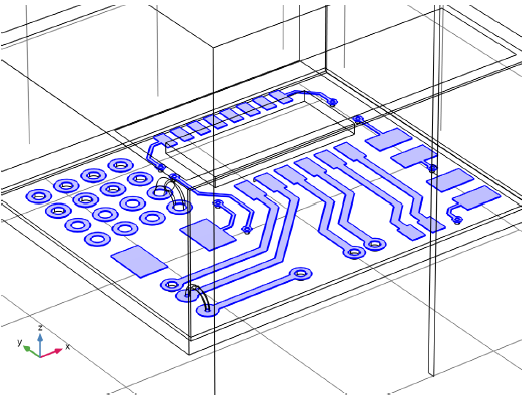}\label{pp3}}
	\caption{The simulation model in COMSOL.}
        \label{simulationsetup}
	\end{figure}

 \begin{figure}[tb]
	\centering
	\subfigure[Electromagnetic field distribution.]{\includegraphics[width=3.3cm]{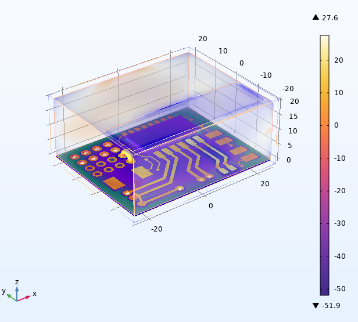}\label{pp4}}
	\subfigure[Shielding effectiveness.]{\includegraphics[width=4.7cm]{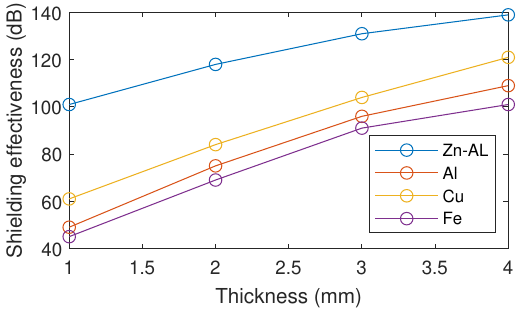}\label{pp5}}
	\caption{Simulation result of the shielding effectiveness under different thicknesses.}
         \label{simulationresult}
	\end{figure}

\subsection{The Comparision with Previous EMI Works}
\label{Comparision}
\subsubsection{Comparision with previous EMI works}
The detailed comparison of EMI attack works is shown in \Cref{comparision}, and the comparison of EMI defense works is shown in \Cref{defense}.
\subsubsection{Comparision with HallSpoofing}
HallSpoofing~\cite{barua2020hall} reveals the threat of static electromagnetic field on Hall current sensors. The distinctions between Hallspoofing and our work are as follows: 
\ding{172} Hallspoofing is limited to manipulating Hall current sensors, whereas our work %possesses 
addresses the threat of EMI on both Hall and non-Hall sensors. Notably, non-Hall sensors in inverters 
may render Hallspoofing impractical for precise manipulations. %precisely. 
\ding{173} Due to the constraints of the magnetic field, the attack distance of Hallspoofing is restricted to a few centimeters. 
\ding{174} In contrast to Hallspoofing, our analysis is comprehensive, delving into vulnerabilities within the inverter's control algorithms.
\ding{175} We revealed a previously unrecognized threat that can directly result in irreversible physical damage to the inverter. Note that achieving \damage involves targeting the DC bus voltage sensor, which is distinct from Hall current sensors.

\subsection{Simulation of Shielding Thickness}
\label{simu}
In the simulation, we set up a dipole antenna as the EMI transmitting source at a distance of 10 mm from the metal casing, with a signal amplitude of 1 V and a frequency of 1000 MHz, and we measured the EMI shielding effectiveness of four kinds of metals.

The result is shown in Fig.~\ref{simulationresult}. As we can see: \ding{172}~in ideal conditions, the 2 mm thickness of various metals can meet the electromagnetic shielding requirements of the military and aviation field; \ding{173}~higher thickness can achieve better shielding effect; \ding{174}~alloy metal shielding effectiveness is higher than pure metal, which is also often used in engineering. Therefore, under the assumption that the inverter case has adopted the optimal material solution, we recommend that inverter manufacturers increase the thickness of the case metal after weighing the heat dissipation, weight, and other indicators.

\end{document}